\documentclass[pra,reprint,twocolumn,showpacs,preprintnumbers,amsmath,amssymb,amsfonts,superscriptaddress]{revtex4}

\usepackage{graphicx}% Include figure files
\usepackage{dcolumn}% Align table columns on decimal point
\usepackage{bm}% bold math
\usepackage{color}% bold math
\usepackage{amsmath,amsthm,amssymb,bm}
\usepackage{mathrsfs}
\usepackage{comment}

\newcommand{\Rb}{^{\text{87}}\text{Rb}}

\newcommand{\Hamil}[1]{\hat{\mathcal{H}}_{#1}}

\newcommand{\annih}[1]{\hat{\psi}_{#1}}
\newcommand{\creat}[1]{\hat{\psi}^\dagger_{#1}}
\newcommand{\ann}[1]{\hat{\delta}_{#1}}
\newcommand{\cre}[1]{\hat{\delta}^\dagger_{#1}}
\newcommand{\lamb}[1]{\hat{\Lambda}_{#1}}

\newcommand{\ahat}{\hat{a}}
\newcommand{\adagg}{\hat{a}^\dagger}

\newcommand{\R}{\mathbf{r}}
\newcommand{\K}{\mathbf{k}}
\newcommand{\s}{\mathbf{s}}

\newcommand{\integral}{\int\text{d}\mathbf{r}}
\newcommand{\integrals}{\int\text{d}\mathbf{s}}

\newcommand{\roundt}{\frac{\partial}{\partial t}}

\newcommand{\order}[1]{^{(#1)}}

\begin{document}

\preprint{APS/123-QED}

%\begin{minipage}{18cm}

\title{Effects of thermal and quantum fluctuations on the phase diagram of a spin-1 $\Rb$ Bose-Einstein condensate}
\author{Nguyen Thanh Phuc}
\affiliation{Department of Physics, University of Tokyo, 7-3-1 Hongo, Bunkyo-ku, Tokyo 113-0033, Japan}
\author{Yuki Kawaguchi}
\affiliation{Department of Physics, University of Tokyo, 7-3-1 Hongo, Bunkyo-ku, Tokyo 113-0033, Japan}
\author{Masahito Ueda}
\affiliation{Department of Physics, University of Tokyo, 7-3-1 Hongo, Bunkyo-ku, Tokyo 113-0033, Japan}
\affiliation{ERATO Macroscopic Quantum Control Project, 7-3-1 Hongo, Bunkyo-ku, Tokyo 113-0033, Japan}
\date{\today}

%%%%%%%%%%%%%%%%%%%%%%%%%%
\begin{abstract}
We investigate effects of thermal and quantum fluctuations on the phase diagram of a spin-1 $\Rb$ Bose-Einstein condensate (BEC) under a quadratic Zeeman effect. Due to the large ratio of spin-independent to spin-dependent interactions of $\Rb$ atoms, the effect of noncondensed atoms on the condensate is much more significant than that in scalar BECs. We find that the condensate and spontaneous magnetization emerge at different temperatures when the ground state is in the broken-axisymmetry phase. In this phase, a magnetized condensate induces spin coherence of noncondensed atoms in different magnetic sublevels, resulting in temperature-dependent magnetization of the noncondensate. We also examine the effect of quantum fluctuations on the order parameter at absolute zero, and find that the ground-state phase diagram is significantly altered by quantum depletion.
\end{abstract}

% PACS, the Physics and Astronomy
                             % Classification Scheme.
%03.75.Hh Static properties of condensates; thermodynamical, statistical, and structural properties 
%03.75.Mn Multicomponent condensates; spinor condensates 
%67.85.Jk Other Bose-Einstein condensation phenomena 
%03.75.Kk Dynamic properties of condensates; collective and hydrodynamic excitations, superfluid flow
%67.30.he Textures and vortices

%\keywords{keywords}%Use showkeys class option if keyword
                              %display desired
\maketitle
%\end{minipage}
%\tableofcontents

%%%%%%%%%%%%%%%%%%%%%%%%%%
\section{Introduction}
\label{sec:Introduction}
Since the first experimental realization of a Bose-Einstein condensate with spin degrees of freedom (spinor BEC) in 1998~\cite{Stamper-Kurn98,Stenger98}, many interesting phenomena have been investigated. Due to the competition between the interatomic interactions and the coupling of atoms to an external magnetic field~\cite{Ho98,Ohmi98}, these systems can exhibit various phases having different spinor order parameters~\cite{Stenger98}. Both theoretical and experimental studies have extensively been conducted on various aspects of spinor BECs (see, for example,~\cite{Ueda10}). Experiments have been performed to investigate formation of spin domains~\cite{Miesner99} or tunneling between them~\cite{Stamper-Kurn99}. Spin-mixing dynamics has also been observed in both spin-1 and spin-2 BECs~\cite{Chang04,Schmaljohann04,Kuwamoto04,Chang05,Kronjager05}. More recently, precise control of the magnetic field has enabled experimenters to observe amplification of spin fluctuations~\cite{Leslie09,Klempt10} and real-time dynamics of spin vortices and short-range spin textures~\cite{Sadler06,Vengalattore08,Vengalattore10}. Finite-temperature properties of spinor BECs have also been theoretically investigated: the dynamics of spinor systems in quasi-one~\cite{Petit06,Cardoner07} and three-dimensional spaces~\cite{Gawryluk07}, and finite-temperature phase diagrams of both ferromagnetic~\cite{Isoshima00,Zhang04,Huang02} and antiferromagnetic spinor condensates~\cite{Szabo07,Mukerjee06,James11}. 

For scalar BECs, the first-order self-consistent approximation (also called the Popov approximation~\cite{Griffin96}), which neglects the pair correlation of noncondensed atoms or the anomalous average, can give a good description of thermal equilibrium properties of the system over a wide range of temperatures except near the BEC transition point. This is because at temperatures well above absolute zero, the anomalous average is negligibly small compared with the noncondensate number density. In contrast, near absolute zero the anomalous average is of the same order of magnitude as the noncondensate number density, but both of them are very small compared with the condensate density and hence negligible. However, for spinor BECs, in particular, spin-1~$\Rb$ BECs, due to the large ratio of spin-independent to spin-dependent interactions, the anomalous average and noncondensate number density are expected to be as important as the spin-dependent interaction between two condensed atoms near absolute zero.

The above striking difference between the scalar and spinor BECs has hitherto not been fully studied. A full investigation of this problem is the main theme of this paper. In Refs.~\cite{Isoshima00,Zhang04}, the quadratic Zeeman energy, which is a key control parameter in spinor BECs, was not taken into account. In the present theoretical study, we investigate effects of thermal and quantum fluctuations in a spinor Bose gas in the presence of the quadratic Zeeman effect. We consider a three-dimensional uniform system of spin-1~$\Rb$ atoms with a ferromagnetic interaction, where the spin-independent interaction is stronger than the spin-dependent interaction by a factor of about 200. Therefore, even when the fraction of noncondensed atoms is small, they can significantly affect the magnetism of the system via the spin-independent interaction.

In this paper, we first use the first-order self-consistent approximation to obtain the finite-temperature phase diagram in the presence of a quadratic Zeeman effect. We find that the system undergoes a two-step phase transition, where condensation and spontaneous magnetization occur at different temperatures. We then examine temperature-dependent magnetization of the noncondensate, which is a remarkable consequence of the spin coherence induced by the magnetized condensate. To investigate the effect of quantum depletion on the phase diagram at absolute zero, we adopt the method developed in Ref.~\cite{Castin98}, in which the order parameter is expanded in powers of the square root of the noncondensate fraction. By applying the method to spinor systems, we find a significant modification of the ground-state phase diagram due to the effect of noncondensed atoms.

This paper is organized as follows. Section~\ref{sec:Hamiltonian and mean-field ground state} introduces a theoretical framework of spin-1 spinor BECs, and describes the mean-field ground-state phase diagram analytically. Section~\ref{sec:Finite-temperature phase diagram under the Popov approximation} discusses the finite-temperature phase diagram by using the first-order self-consistent approximation, and studies magnetizations of the condensate and noncondensate as functions of temperature. Section~\ref{sec:Effects of noncondensed Atoms on the Ground State at Absolute Zero} investigates the effect of quantum depletion on the zero-temperature ground-state phase diagram. The perturbative expansion method for spinor BECs is introduced, followed by a discussion of a modification of the ground-state phase diagram from the first-order counterpart. Finally, Sec.~\ref{sec:conclusions} concludes this paper by discussing possible experimental situations. Complicated algebraic manipulations that would distract readers from the main subject are placed in Appendices.

%%%%%%%%%%%%%%%%%%%%%%%%%%
\section{Hamiltonian and mean-field ground state}
\label{sec:Hamiltonian and mean-field ground state}
We consider a system of spin-1 identical bosons with mass $M$ that are confined by an external potential $U(\R)$ and subject to a magnetic field in the $z$~direction. The one-body part of the Hamiltonian is given in matrix form by
\begin{align}
(h_0)_{ij}= \left[-\frac{\hbar^2\nabla^2}{2M}+U(\R)-pi+qi^2\right]\delta_{ij},
\label{eq:h0}
\end{align}
where the subscripts $i,j=0,\pm 1$ refer to the magnetic sublevels, and $p$ and $q$ are the coefficients of the linear and quadratic Zeeman terms, respectively. The total Hamiltonian of the spin-1 spinor Bose gas is given in the second quantization by~\cite{Ho98,Ohmi98}
\begin{align}
\Hamil{}=&\integral\sum_{i,j}\Bigg[ \creat{i}(\R)(h_0)_{ij} \annih{j}(\R)\nonumber\\
&+\frac{c_0}{2} \creat{i}(\R)\creat{j}(\R)\annih{j}(\R)\annih{i}(\R)\Bigg] \nonumber\\
&+\frac{c_1}{2}\sum\limits_{\alpha,i,j,k,l} (f_\alpha)_{ij}(f_\alpha)_{kl}\creat{i}(\R)\creat{k}(\R)\annih{l}(\R)\annih{j}(\R),
\label{eq: Spin-1 spinor BEC's Hamiltonian}
\end{align}
where $\annih{i}(\R)$ is the field operator that annihilates an atom in the magnetic sublevel $i$ at position $\R$, $\alpha=x,y$, or $z$ specifies the spin component, and $f_\alpha$'s denote the components of the spin-1 matrix vector given by
\begin{align}
f_x=&\frac{1}{\sqrt{2}}
   \begin{pmatrix}
   0&1&0\\
   1&0&1\\
   0&1&0\\
   \end{pmatrix},\\
f_y=&\frac{i}{\sqrt{2}}
   \begin{pmatrix}
   0&-1&0\\
   1&0&-1\\
   0&1&0\\
   \end{pmatrix},\\
f_z=&
   \begin{pmatrix}
   1&0&0\\
   0&0&0\\
   0&0&-1\\
   \end{pmatrix}.
\end{align}
The last two terms in the Hamiltonian~(\ref{eq: Spin-1 spinor BEC's Hamiltonian}) describe the spin-independent and spin-dependent interactions, respectively. The coefficients $c_0$ and $c_1$ can be expressed in terms of the $s$-wave scattering lengths $a_0$ and $a_2$ of binary collisions with total spin $F_\mathrm{total}=0$ and $2$, respectively, as~\cite{Ho98}
\begin{subequations}
\begin{align}
c_0=&\frac{4\pi\hbar^2}{M}\frac{a_0+2a_2}{3},\\
c_1=&\frac{4\pi\hbar^2}{M}\frac{a_2-a_0}{3}.
\end{align}
\label{eq: Definition of c0,c1}
\end{subequations}

In the mean-field ground state of a spinor Bose gas, the effect of quantum depletion is neglected, and all particles are assumed to occupy the same single-particle state in both coordinate and spin spaces. The field operator $\annih{i}(\R)$ can then be replaced by a classical field $\phi_i(\R)$, and the expectation value of Hamiltonian~(\ref{eq: Spin-1 spinor BEC's Hamiltonian}) is given by the following energy functional:
\begin{align}
E[\phi_i]=\integral \left[ \sum\limits_{i,j}\phi_i^*(h_0)_{ij}\phi_j+\frac{c_0}{2}(n^\mathrm{c})^2+\frac{c_1}{2}|\mathbf{F}^\mathrm{c}|^2 \right],
\label{eq:energy functional}
\end{align}
where the number density $n^\mathrm{c}(\R)$ and the three components of the spin density vector $\mathbf{F}^\mathrm{c}(\R)$ of the condensate are given by
\begin{align}
n^\mathrm{c} \equiv& \sum\limits_i |\phi_i(\R)|^2, 
\label{eq:nc}\\
F^\mathrm{c}_\alpha(\R) \equiv& \sum\limits_{i,j} \phi_i^*(\R)(f_\alpha)_{ij}\phi_j(\R)\qquad (\alpha=x,y,z).
\label{eq:Fc}
\end{align}
In the mean-field approximation, $n^\mathrm{c}$ is equal to the total number density $n$. The condensate wave function $\phi_i(\R)$ is determined by minimizing the energy functional~(\ref{eq:energy functional}), i.e.,
\begin{align}
\frac{\delta E[\phi_i]}{\delta\phi_i^*(\R)}=0,
\label{eq: variation of energy functional}
\end{align}
subject to the normalization condition
\begin{align}
\integral \sum\limits_i |\phi_i(\R)|^2=N,
\label{eq: normalization condition 1}
\end{align}
where $N$ is the total number of atoms.
Equation~\eqref{eq: variation of energy functional}, together with Eq.~\eqref{eq: normalization condition 1}, leads to the multi-component Gross-Pitaevskii (GP) equation:
\begin{align}
\sum_{j}\left[(h_0)_{ij}+c_0n\delta_{ij}+c_1\sum\limits_{\alpha}F^\mathrm{c}_\alpha(f_\alpha)_{ij}\right]\phi_j=\mu\phi_i,
\label{eq:GP equation}
\end{align}
where $\mu$ is the chemical potential at absolute zero.

\begin{figure}[tbp] % float placement: (h)ere, page (t)op, page (b)ottom, other (p)age
  \centering
  % file name: F:/Paper Drafts (Apr 11th, 2011)/Figures/Fzx_mfgroundstate.eps
  \includegraphics[bb=0 0 258 245,width=3in,height=2.85in,keepaspectratio]{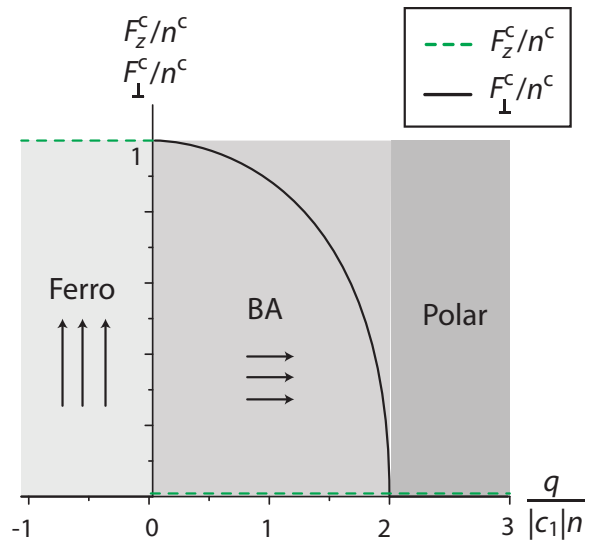}
  \caption{(Color online) $q$ dependence of longitudinal magnetization per particle $F^\mathrm{c}_z/n^\mathrm{c}$ and that of transverse one $F^\mathrm{c}_\perp/n^\mathrm{c}$ in the mean-field ground state of a spin-1 ferromagnetic BEC ($c_1<0$). The dashed line and solid curve show the longitudinal and transverse magnetizations, respectively, where $F^{\rm c}_\perp\equiv |F^{\rm c}_+|=\sqrt{(F^{\rm c}_x)^2+(F^{\rm c}_y)^2}$. Ferro, BA and Polar stand for ferromagnetic, broken-axisymmetry, and polar phases, which are shaded in light, medium and dark grey, respectively. The longitudinal magnetization per particle is $F^\mathrm{c}_z/n^\mathrm{c}=\Theta(-q)$, and the transverse one is $|F^\mathrm{c}_+|/n^\mathrm{c}=\sqrt{1-(q/2|c_1|n)^2}\Theta(q)\Theta(2|c_1|n-q)$, where $\Theta(q)$ is the unit-step function.}
  \label{fig:mean-field ground-state FzFx}
\end{figure}

For a uniform system, i.e., when $U(\R)=0$, the condensate wave function $\phi_i$ is independent of $\R$ and the solutions to Eq.~(\ref{eq:GP equation}) can be obtained analytically. For the case of $c_1<0$ and $p=0$, which is the case we consider in the present paper, the order parameters $\bm{\phi}=(\phi_1,\phi_0,\phi_{-1})^\mathrm{T}$ and the energies per particle $\epsilon=E[\phi_i]/N$ for possible phases are given as follows~\cite{Stenger98,Murata07}:
\begin{align}
{\rm Ferro:\ }
&\bm{\phi} = \sqrt{n}(1,0,0)^{\rm T}
\ \ \textrm{or}\ \ \sqrt{n}(0,0,1)^{\rm T}, \label{eq: ferro phase order parameter}\\
&\epsilon=q+\frac{c_0+c_1}{2}n,\\
{\rm Polar:\ }
&\bm{\phi} =  \sqrt{n} (0,1,0)^{\rm T}, \label{eq: polar phase order parameter}\\
&\epsilon = \frac{c_0}{2}n,\\
{\rm BA:\ }
& \bm{\phi} = \sqrt{\frac{n}{2}}
\begin{pmatrix}
e^{-i\theta}\sqrt{\frac{1}{2}\left(1-\frac{q}{2|c_1|n}\right)}\\
\sqrt{ 1+\frac{q}{2|c_1|n}}\\
e^{i\theta}\sqrt{\frac{1}{2}\left(1-\frac{q}{2|c_1|n}\right)}\\
\end{pmatrix}, \label{eq: BA phase order parameter}\\
&\epsilon=\left(1-\frac{q}{2|c_1|n}\right)^2\frac{c_1}{2}n
+\frac{c_0}{2}n,
\end{align}
where $\mathrm{T}$ denotes transpose, and Ferro, Polar, and BA stand for ferromagnetic, polar, and broken-axisymmetry phases, respectively. In Eq.~\eqref{eq: BA phase order parameter}, $\theta$ can take on values between $0$ and $2\pi$, and we have omitted overall phase factors in Eqs.~\eqref{eq: ferro phase order parameter},~\eqref{eq: polar phase order parameter}, and~\eqref{eq: BA phase order parameter}. The BA phase exists only in the region of $0<q<2|c_1|n$, and becomes the ground state of the system in this  parameter regime. The magnetization for the BA phase is transverse and given by
\begin{align}
F^{\rm c}_z &=0,\\
F^{\rm c}_+&\equiv F^{\rm c}_x+iF^{\rm c}_y = n e^{i\theta} \sqrt{1-\left(\frac{q}{2c_1n}\right)^2}.
\end{align}
Hence, $\theta$ specifies the direction of magnetization in the $xy$-plane, and its magnitude depends on $q$. The BA phase is named after the fact that the transverse magnetization breaks the rotational symmetry of the Hamiltonian around the $z$-axis~\cite{Murata07}. If $q>2|c_1|n$, the ground state is in the polar phase. In this phase, the condensate has zero magnetization. On the other hand, if $q<0$, the fully polarized state in the magnetic sublevel $i=1$ or $-1$ minimizes both the ferromagnetic interaction and the quadratic Zeeman energy. Therefore, the ferromagnetic phase is the ground state of the system. To satisfy the conservation of the total longitudinal magnetization, a phase separation into two spin domains with $F^{\rm c}_z/n^{\rm c}=1$ and $-1$ must occur. Figure~\ref{fig:mean-field ground-state FzFx} shows the $q$ dependence of the longitudinal magnetization and that of the transverse one.

%%%%%%%%%%%%%%%%%%%%%%%%%%
\section{Finite-temperature phase diagram under the first-order self-consistent approximation}
\label{sec:Finite-temperature phase diagram under the Popov approximation}
\subsection{First-order self-consistent approximation}
\label{subsec:Popov approximation}
At finite temperatures, a fraction of atoms are thermally excited from the condensate to form a thermal cloud, which, in turn, will affect the condensate. Therefore, the finite-temperature phase diagram should be determined in a self-consistent manner. The field operator is decomposed into the condensate part, which can be replaced by a classical field $\phi_i(\R)$, and the noncondensate part $\ann{i}(\R)$:
\begin{equation}
\annih{i}(\R)=\phi_i(\R)+\ann{i}(\R).
\label{eq: field operator in spinor BECs}
\end{equation}
For convenience, we consider here a grand-canonical ensemble of the atomic system and introduce the operator
\begin{align}
\hat{\mathcal{K}}=\Hamil{}-\mu\hat{\mathcal{N}},
\end{align}
where the total number operator $\hat{\mathcal{N}}$ is defined as
\begin{align}
\hat{\mathcal{N}}\equiv\integral\sum_i\creat{i}(\R)\annih{i}(\R).
\label{eq:N}
\end{align}
Substituting Eq.~(\ref{eq: field operator in spinor BECs}) into Eqs.~(\ref{eq: Spin-1 spinor BEC's Hamiltonian}) and \eqref{eq:N}, and collecting terms of the same order with respect to the fluctuation operator $\ann{i}(\R)$, we obtain
\begin{equation}
\hat{\mathcal{K}}=K_0+\hat{K}_1+\hat{K}_2+\hat{K}_3+\hat{K}_4,
\label{eq: H=H0+H1+...+H4}
\end{equation}
where $\hat{K}_n\;(n=0,\dots,4)$ is comprised of the terms that involve the $n$-th power of $\ann{i}(\R)$.

The static properties of the system in thermal equilibrium can be calculated from the eigenspectrum of operator $\hat{\mathcal{K}}$. The part of $\hat{\mathcal{K}}$ that involves the terms up to quadratic in $\ann{i}$, i.e., $K_0+\hat{K}_1+\hat{K}_2$, can be diagonalized using a Bogoliubov transformation~\cite{Bogoliubov47}. The higher-order terms in $\hat{K}_3$ and $\hat{K}_4$ can be made into quadratic forms by applying the mean-field approximation to noncondensed atoms. The mean-field approximation for noncondensate operators in $\hat{K}_3$ and $\hat{K}_4$ is carried out as follows~\cite{Griffin96,Proukakis08}:
\begin{align}
\cre{i}\ann{j}\ann{k} \simeq \ & \tilde{n}_{ij}\ann{k}+\tilde{n}_{ik}\ann{j}+\cre{i}\tilde{m}_{jk},\\
\cre{i}\cre{j}\ann{k}\ann{l} \simeq \ &\tilde{n}_{ik}\cre{j}\ann{l}+\tilde{n}_{jl}\cre{i}\ann{k}-\tilde{n}_{ik}\tilde{n}_{jl}\nonumber\\
&+\tilde{n}_{il}\cre{j}\ann{k}+\tilde{n}_{jk}\cre{i}\ann{l}-\tilde{n}_{il}\tilde{n}_{jk}\nonumber\\
&+\tilde{m}^*_{ij}\ann{k}\ann{l}+\tilde{m}_{kl}\cre{i}\cre{j}-\tilde{m}^*_{ij}\tilde{m}_{kl},
\label{eq: mean-field approximation for noncondensed atoms}
\end{align}
where $\tilde{n}_{ij}(\R) \equiv \langle \cre{i}(\R)\ann{j}(\R) \rangle$ is the matrix element of the noncondensate number density, and $\tilde{m}_{ij}(\R) \equiv \langle \ann{i}(\R)\ann{j}(\R) \rangle$ is that of the noncondensate pair correlation, which is called the anomalous average.

In the first-order self-consistent approximation, the anomalous averages $\tilde{m}_{ij}(\R)$ and $\tilde{m}_{ij}^*(\R)$ are neglected~\cite{Griffin96}. This would give a gapless spectrum of elementary excitations, in agreement with the Nambu-Goldstone theorem. At temperatures well above absolute zero, the spectrum of elementary excitations approaches that of single particles, and, therefore, the anomalous average $\tilde{m}_{ij}$ becomes negligibly small compared with the noncondensate number density $\tilde{n}_{ij}$ \cite{Yukalov05}. Consequently, the first-order self-consistent approximation gives a good description of a Bose gas in thermal equilibrium over a broad range of  temperatures, except near absolute zero.

The condensate wave function $\phi_i(\R)$ then satisfies the generalized GP equation, which is obtained from the requirement that the operator $\hat{\mathcal{K}}$ be stationary with respect to $\phi_i(\R)$, or equivalently, that the sum of terms that are linear in $\ann{i}(\R)$ vanish:
\begin{align}
&\sum\limits_j\Bigg\{(h_0)_{ij}\phi_j+c_0(n^\mathrm{c}+n^\mathrm{nc})\delta_{ij}+c_0\tilde{n}_{ij}^*\phi_j \nonumber \\
&+c_1\sum\limits_{\alpha}\Bigg[(F^\mathrm{c}_\alpha+F^\mathrm{nc}_\alpha)(f_\alpha)_{ij}+\sum\limits_{k,l}(f_\alpha)_{ik}(f_\alpha)_{lj}\tilde{n}_{kl}^*\Bigg]\Bigg\}\phi_j\nonumber\\
&=\mu\phi_i,
\label{eq: generalized GP eq. for spinor BEC in Popov approximation}
\end{align}
where the condensate number density $n^{\rm c}(\R)$ and spin density $F^{\rm c}_\alpha(\R)$ ($\alpha=x,y,z$) are defined in Eqs.~(\ref{eq:nc}) and (\ref{eq:Fc}), respectively, and $n^{\rm nc}(\R)$ and $F^{\rm nc}_\alpha(\R)$ are the noncondensate counterparts given by
\begin{align}
n^\mathrm{nc}(\R) \equiv& \sum\limits_i \tilde{n}_{ii}(\R), \label{eq:nt} \\
F^\mathrm{nc}_\alpha(\R) \equiv& \sum\limits_{i,j}(f_\alpha)_{ij}\tilde{n}_{ij}(\R).
\label{eq:nc,nt,Fc,Ft}
\end{align}
Thus, the operator $\hat{\mathcal{K}}$ reduces to the sum of a c-number term$K_0$ and a quadratic operator $\hat{\mathcal{K}}^{(2)}$, where
\begin{align}
K_0=&\integral \Bigg[ \sum\limits_{i,j}\phi_i^*(h_0)_{ij}\phi_j-\mu n^\mathrm{c}\nonumber\\
&+\frac{c_0}{2}(n^\mathrm{c})^2+\frac{c_1}{2}|\mathbf{F}^\mathrm{c}|^2 \Bigg],\\
\hat{\mathcal{K}}^{(2)}=&\integral \sum\limits_{i,j}\Bigg[\cre{i}A_{ij}(\R)\ann{j}+\frac{1}{2}\Bigg(\cre{i}B_{ij}(\R)\cre{j}\nonumber\\
&+\ann{i}B_{ij}^*(\R)\ann{j}\Bigg)\Bigg].
\end{align}
Here, the matrices $A_{ij}(\R)$ and $B_{ij}(\R)$ are defined as
\begin{subequations}
\label{eq: A,B matrices}
\begin{align}
A_{ij}(\R)\equiv\ &(h_0)_{ij}-\mu\delta_{ij}+c_0\Big[(n^\mathrm{c}+n^\mathrm{nc})\delta_{ij}\nonumber\\
&+(\phi_i\phi_j^*+\tilde{n}_{ij}^*)\Big]+c_1\sum\limits_\alpha\Bigg[(F^\mathrm{c}_\alpha+F^\mathrm{nc}_\alpha)(f_\alpha)_{ij}\nonumber\\
&+\sum\limits_{k,l}(f_\alpha)_{il}(f_\alpha)_{kj}(\phi_k^*\phi_l+\tilde{n}_{kl})\Bigg],\\
B_{ij}(\R)\equiv\ &c_0\phi_i\phi_j+c_1\sum\limits_{\alpha,k,l}(f_\alpha)_{ik}(f_\alpha)_{jl}\phi_k\phi_l.
\end{align}
\end{subequations}
We diagonalize the quadratic operator $\hat{\mathcal{K}}^{(2)}$ by a Bogoliubov transformation
\begin{equation}
\hat{b}^{(\lambda)}=\integral \sum_i \Big[u_i^{(\lambda)*}(\R)\ann{i}(\R)-v_i^{(\lambda)}(\R)\cre{i}(\R)\Big],
\end{equation}
where the coefficients $u_i^{(\lambda)}(\R)$ and $v_i^{(\lambda)}(\R)$ ($i=0,\pm 1$) satisfy the generalized Bogoliubov-de Gennes (BdG) equation for the excitation mode labeled by index $\lambda$:
\begin{equation}
\begin{pmatrix}
   A_{ij}(\R) & B_{ij}(\R) \\
   -B_{ij}^*(\R) & -A_{ij}^*(\R)\\
   \end{pmatrix}
   \begin{pmatrix}
   u^{(\lambda)}_j(\R)\\
   v^{(\lambda)}_j(\R)\\
   \end{pmatrix}
   =\epsilon^{(\lambda)}
   \begin{pmatrix}
   u^{(\lambda)}_i(\R)\\
   v^{(\lambda)}_i(\R)\\
   \end{pmatrix}.
\label{eq: BdG equation for spinor BEC}
\end{equation}
In thermal equilibrium, the noncondensate number density is expressed in terms of $u^{(\lambda)}(\R)$ and $v^{(\lambda)}(\R)$ as
\begin{align}
\tilde{n}_{ij}(\R)=&\sum_\lambda\Big\{u_i^{(\lambda)*}(\R) u_j^{(\lambda)}(\R)f(\epsilon^{(\lambda)})\nonumber\\
&+v_i^{(\lambda)}(\R) v_j^{(\lambda)*}(\R)\left[f(\epsilon^{(\lambda)})+1\right] \Big\},
\label{eq:nij}
\end{align}
where $f(\epsilon)=1/[\exp(\epsilon/k_BT)-1]$ is the Bose-Einstein distribution function, and the coefficients $u^{(\lambda)}(\R)$ and $v^{(\lambda)}(\R)$ are normalized as
\begin{align}
\integral \sum\limits_i \Big[|u_i^{(\lambda)}(\R)|^2-|v_i^{(\lambda)}(\R)|^2\Big]=1.
\end{align}
Finally, the condensate and noncondensate satisfy the following number equation:
\begin{equation}
N=\integral \left[n^\mathrm{c}(\R)+n^\mathrm{nc}(\R)\right].
\label{eq: number equation for spinor BECs}
\end{equation}

%%%%%%%%%%%%%%%%%%%%%%%%%%
\subsection{Finite-temperature phase diagram}
\label{subsec:Finite-temperature Phase Diagram}
In the following sections, we consider a three-dimensional uniform system of spin-1 $\Rb$ atoms with a fixed total number density $n$. Then, the condensate wave function $\phi_i$ and the normal density $\tilde{n}_{ij}$ are constant, while the coefficients of the Bogoliubov transformation are given by
\begin{subequations}
\label{eq:plane waves}
\begin{align}
u^{(\lambda)}_j(\R) &= u^{(\nu,\K)}_j e^{i \K \cdot \R},\\
v^{(\lambda)}_j(\R) &= v^{(\nu,\K)}_j e^{i \K \cdot \R},
\end{align}
\end{subequations}
where $\K$ is the wave vector and $\nu$ is an index to distinguish between excitation modes.

We consider the case in which the system is initially prepared so that the total magnetization projected along the $z$-axis vanishes. Due to the conservation of the total longitudinal magnetization, the linear Zeeman term vanishes, and, therefore, we have $p=0,q\not=0$ in Eq.~\eqref{eq:h0}. The $s$-wave scattering lengths of the $^{87}$Rb atom in the $F=1$ hyperfine manifold are calculated to be $a_0=101.8\, a_{\mathrm{B}}$ and $a_2=100.4\, a_{\mathrm{B}}$~\cite{Klausen01}, where $a_\mathrm{B}$ is the Bohr radius. Consequently, $c_1$ given in Eq.~(\ref{eq: Definition of c0,c1}) is negative, i.e., the interaction is ferromagnetic, and it is about 200 times smaller than $c_0$.

We have numerically solved a set of coupled equations in the first-order self-consistent approximation [Eqs.~(\ref{eq: generalized GP eq. for spinor BEC in Popov approximation})--(\ref{eq: number equation for spinor BECs})] at a given temperature and for a given value of $q$. Here, the generalized GP equation (\ref{eq: generalized GP eq. for spinor BEC in Popov approximation}) was solved numerically by using the imaginary-time propagation method, which evolves a randomly chosen initial state to a local minimum of the Hamiltonian.

\begin{figure}[tbp] % float placement: (h)ere, page (t)op, page (b)ottom, other (p)age
  \centering
  % file name: F:/Paper Drafts (May 20th, 2011)/Figures/finiteTdiagram.EPS
  \includegraphics[bb=0 0 258 464,width=3in,height=5.4in,keepaspectratio]{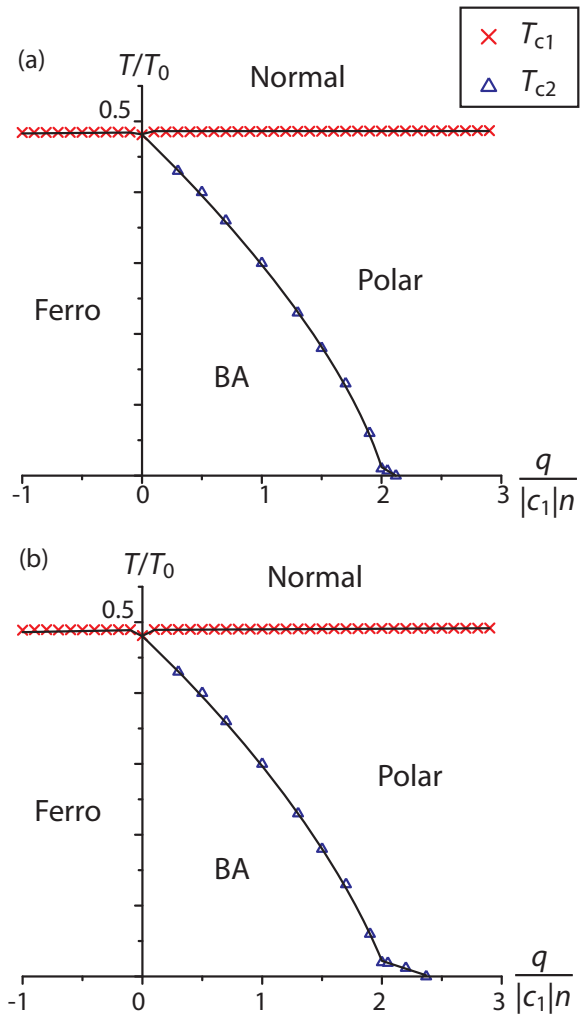}
  \caption{(Color online) Finite-temperature phase diagram of a spin-1 ferromagnetic Bose gas in the first-order self-consistent approximation. We have used the interaction parameters for $\Rb$ atoms, i.e., $a_0=101.8~a_{\mathrm{B}}$ and $a_2=100.4~a_{\mathrm{B}}$~\cite{Klausen01}, with the total number density (a) $n=1.0\times 10^{12}~\mathrm{cm}^{-3}$, and (b) $n=1.0\times 10^{13}~\mathrm{cm}^{-3}$. The quadratic Zeeman energy $q$ and temperature $T$ are measured in units of the spin-dependent interaction $|c_1|n$ and the transition temperature of a uniform ideal scalar BEC, $T_0=3.31\hbar^2n^{2/3}/(k_\mathrm{B}M)$, respectively. Crosses show temperature $T_{\mathrm{c}1}$ below which the condensate density $n^\mathrm{c}$ becomes nonzero. Open triangles show temperature $T_{\mathrm{c}2}$ below which the condensate acquires a nonzero transverse magnetization $F^\mathrm{c}_\perp$. The solid curves show guides for the eye.}
  \label{fig: Phase Diagram of Rb at finite temperature under quadratic Zeeman effect}
\end{figure}

Figure~\ref{fig: Phase Diagram of Rb at finite temperature under quadratic Zeeman effect} shows the finite-temperature phase diagram of a spin-1 $\Rb$ BEC with (a) $n=1.0\times 10^{12}~\mathrm{cm}^{-3}$ and (b) $n=1.0\times 10^{13}~\mathrm{cm}^{-3}$. Here, the phase of the system is identified by calculating the condensate number density $n^\mathrm{c}$ and the longitudinal $F^\mathrm{c}_z$ and transverse $F^\mathrm{c}_{\perp}\equiv \sqrt{(F^\mathrm{c}_x)^2+(F^\mathrm{c}_y)^2}$ magnetizations of the condensate. The high-temperature normal phase has $n^\mathrm{c}=0$, while the condensed phases have $n^\mathrm{c}\neq 0$. The ferromagnetic, BA, and polar phases are characterized by $F^\mathrm{c}_z/n^\mathrm{c}=1$, $F^\mathrm{c}_{\perp}\not=0$, and $F^\mathrm{c}_z=F^\mathrm{c}_{\perp}=0$, respectively.

Figure~\ref{fig: Phase Diagram of Rb at finite temperature under quadratic Zeeman effect} shows that the region of BA phase shrinks with increasing temperature because thermal fluctuations suppress the transverse magnetization. The phase boundary between the BA and polar phases is almost independent of the total number density $n$ except near absolute zero. We note that if the ground state is in the BA phase, the phase transition is a two-step process: first, the system undergoes a phase transition from the normal phase to the polar phase at temperature $T_{\mathrm{c}1}\simeq 0.48~T_0$, where $T_0=3.31\hbar^2n^{2/3}/(k_\mathrm{B}M)$ is the transition temperature of a uniform ideal scalar BEC with the same atomic density $n$. Here, $T_{\mathrm{c}1}$ is smaller than $T_0$ by a factor of about $(1/3)^{2/3}\simeq 0.48$, because above $T_{\mathrm{c}1}$, the population of atoms in each magnetic sublevel is almost equal to $n/3$. The quadratic Zeeman effect causes a further small shift of $T_{\mathrm{c}1}$ from the value $0.48~T_0$ by making a slight population imbalance. A slope of the normal-condensate phase boundary caused by the quadratic Zeeman effect is too small ($\sim 10^{-4}$) to be seen in Fig.~\ref{fig: Phase Diagram of Rb at finite temperature under quadratic Zeeman effect}. At a lower temperature $T_{\mathrm{c}2}$ ($T_{\mathrm{c}2}<T_{\mathrm{c}1}$), the system undergoes a second phase transition to the BA phase having a nonzero transverse magnetization.

The physics of the two-step phase transition can be understood as follows. Due to the positive quadratic Zeeman energy, the magnetic sublevel $i=0$ has a higher population than the levels $i=\pm 1$. Consequently, the system first condenses into the polar phase. If the temperature is further lowered, the other states ($i=\pm 1$) also undergo Bose-Einstein condensation, and the system enters the BA phase by developing transverse magnetization.

We note that the phase diagram shown in Fig.~\ref{fig: Phase Diagram of Rb at finite temperature under quadratic Zeeman effect} does not agree with the mean-field phase diagram, in which the phase boundary between the BA and polar phases at $T=0$ is given by $q=q_\mathrm{b}$ with $q_\mathrm{b}=2|c_1|n$. In the first-order self-consistent approximation, the phase boundary shifts to $q_\mathrm{b}=2.12|c_1|n$ for $n=1.0\times 10^{12}~{\rm cm}^{-3}$ and $q_\mathrm{b}=2.37|c_1|n$ for $n=1.0\times 10^{13}~{\rm cm}^{-3}$ due to quantum fluctuations. The results suggest that a more careful treatment needs to be made for the anomalous average near absolute zero. We shall discuss this point in Sec.~\ref{sec:Effects of noncondensed Atoms on the Ground State at Absolute Zero}.

%%%%%%%%%%%%%%%%%%%%%%%%%%
\subsection{Condensate fraction and magnetization}
\label{subsec:Condensate Fraction and Magnetization}
\begin{figure}[tbp] % float placement: (h)ere, page (t)op, page (b)ottom, other (p)age
  \centering
  % file name: F:/Paper Drafts (May 20th, 2011)/Figures/q1p0_new.EPS
  \includegraphics[bb=0 0 302 315,width=3in,height=3.13in,keepaspectratio]{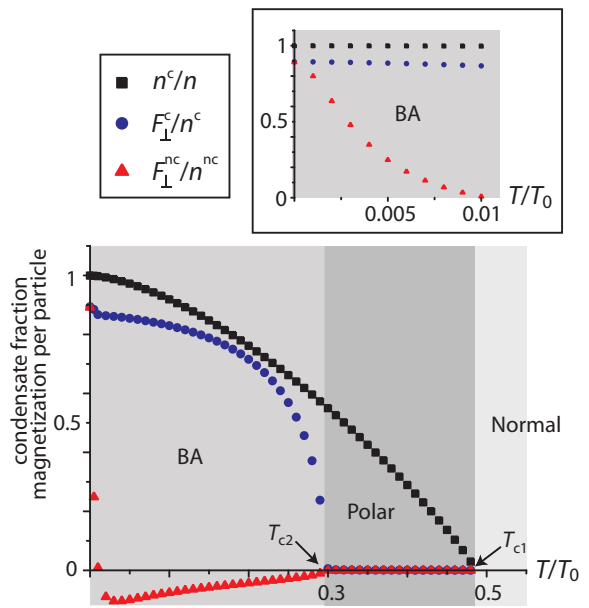}
  \caption{(Color online) Temperature dependence of the condensate fraction $n^\mathrm{c}/n$ (squares), transverse magnetizations per particle of the condensate $F^\mathrm{c}_\perp/n^\mathrm{c}$ (circles) and noncondensate $F^\mathrm{nc}_\perp/n^\mathrm{nc}$ (triangles) for $q=|c_1|n$ and $n=1.0\times 10^{12}~{\rm cm}^{-3}$. The BA, polar, and normal phases are shaded in medium, dark, and light grey, respectively. The inset shows enlarged behaviors of these physical quantities near absolute zero. The longitudinal magnetizations of the condensate and noncondensate vanish for $q>0$; thus, for the BA phase their magnetizations are both perpendicular to the external magnetic field. The negative values of the transverse magnetization of the noncondensate at $0.01~T_0\lesssim T \lesssim 0.3~T_0$ imply that the magnetization vectors of the condensate and noncondensate are antiparallel to each other.}
  \label{fig: Condensate fraction and magnetization for Rb spinor BEC}
\end{figure}

Next, we study the temperature dependence of the condensate fraction $n^\mathrm{c}/n$, and that of the longitudinal and transverse magnetizations per particle of both the condensate $F^\mathrm{c}_{z,\perp}/n^\mathrm{c}$ and noncondensate $F^\mathrm{nc}_{z,\perp}/n^\mathrm{nc}$. Figure~\ref{fig: Condensate fraction and magnetization for Rb spinor BEC} shows the result of numerical calculation for $q=|c_1|n$. For other values of $q$ in the region $0<q<q_\mathrm{b}$, these physical quantities depend on temperature in a manner qualitatively similar to the case of $q=|c_1|n$. The longitudinal magnetizations $F_z^\mathrm{c,nc}$ vanish over the whole temperature region. The transverse magnetizations of the condensate $F^\mathrm{c}_{\perp} \equiv \sqrt{(F^\mathrm{c}_x)^2+(F^\mathrm{c}_y)^2}$ and noncondensate $F^\mathrm{nc}_\perp \equiv \sigma\sqrt{(F^\mathrm{nc}_x)^2+(F^\mathrm{nc}_y)^2}$ are given by
\begin{subequations}
\label{eq:Fc,Ft}
\begin{align}
F^\mathrm{c}_x+iF^\mathrm{c}_y=&\ \sqrt{2}(\phi_1^*\phi_0+\phi_0^*\phi_{-1}),\\
F^\mathrm{nc}_x+iF^\mathrm{nc}_y=&\ \sqrt{2}(\tilde{n}_{1,0}+\tilde{n}_{0,-1}),\label{eq:Ft}
\end{align}
\end{subequations}
where the matrix elements of the noncondensate number density matrix are defined as $\tilde{n}_{ij} \equiv \langle \cre{i}\ann{j} \rangle$ below Eq.~\eqref{eq: mean-field approximation for noncondensed atoms}. The transverse magnetization of the noncondensate is parallel or antiparallel to that of the condensate, and we set $\sigma=1$ ($\sigma=-1$) if they are parallel (antiparallel).

Figure~\ref{fig: Condensate fraction and magnetization for Rb spinor BEC} demonstrates a two-step phase transition, in which a nonzero condensate fraction emerges at temperature $T_{\mathrm{c}1}\simeq 0.48~T_0$, and finite transverse magnetizations of both the condensate and noncondensate emerge at a lower temperature $T_{\mathrm{c}2}\simeq 0.3~T_0$. The nonzero transverse magnetization of the noncondensate is a consequence of the spin coherence of noncondensed atoms, which results from their coupling to the magnetized condensate. Above $T_{\mathrm{c}1}$, where there is no condensate, no spin coherence exists within the thermal cloud. In contrast, above $T_{\mathrm{c}1}$, the condensate induces spin coherence of noncondensed atoms as indicated by $\tilde{n}_{ij}\neq 0$ for $i\neq j$, leading to a nonzero transverse magnetization $F^\mathrm{nc}_\perp$. The spin coherence was experimentally observed in a two-level spinor system at low temperatures~\cite{Lewandowski03,McGuirk03}.

It can also be seen from Fig.~\ref{fig: Condensate fraction and magnetization for Rb spinor BEC} that the magnetization of the condensate and that of the noncondensate are antiparallel to each other over a wide range of temperatures ($0.01\lesssim T/T_0 \lesssim 0.3$) except near absolute zero, where they become parallel to each other. This phenomenon can be understood by considering the energy spectra of the excitation modes of the system described by Eq.~(\ref{eq: BdG equation for spinor BEC}). From Eqs.~(\ref{eq:nij}) and (\ref{eq:Ft}), the transverse magnetization of the noncondensate can be expressed in terms of the excitation modes as
\begin{align}
F^{\rm nc}_+ \equiv\ & F^{\rm nc}_x + i F^{\rm nc}_y\nonumber\\
 = \ &\sum_{\nu} \sum_{\K}\left[F^{\rm tf}_{+,\nu,\K}f(\epsilon^{(\nu,\K)})+ F^{\rm qd}_{+,\nu,\K}\right],
\label{eq:Fthermal}
\end{align}
where $\nu=1,2,3$ denote the index of the excitation modes [see Eq.~(\ref{eq:plane waves})], and
\begin{subequations}
\label{eq:Ftf,Fqd}
\begin{align}
F^{\rm tf}_{+,\nu,\K} =\ & \sqrt{2}\Big[u_1^{(\nu,\K)*}u_0^{(\nu,\K)}+u_0^{(\nu,\K)*}u_{-1}^{(\nu,\K)}\nonumber\\
&+v_1^{(\nu,\K)}v_0^{(\nu,\K)*}+v_0^{(\nu,\K)}v_{-1}^{(\nu,\K)*}\Big],\\
F^{\rm qd}_{+,\nu, \K} =\ & \sqrt{2}
\Big[v_1^{(\nu,\K)}v_0^{(\nu,\K)*}
+v_0^{(\nu,\K)}v_{-1}^{(\nu,\K)*}\Big],
\end{align}
\end{subequations}
give the contribution to $F^{\rm nc}_+$ from the thermally excited collective modes and that from the quantum depletion at absolute zero, respectively. 

\begin{figure}[tbp] % float placement: (h)ere, page (t)op, page (b)ottom, other (p)age
  \centering
  % file name: F:/Paper Drafts (Apr 13th, 2011)/Figures/energy_spectrum.EPS
  \includegraphics[bb=0 0 256 733,width=2.5in,height=7.17in,keepaspectratio]{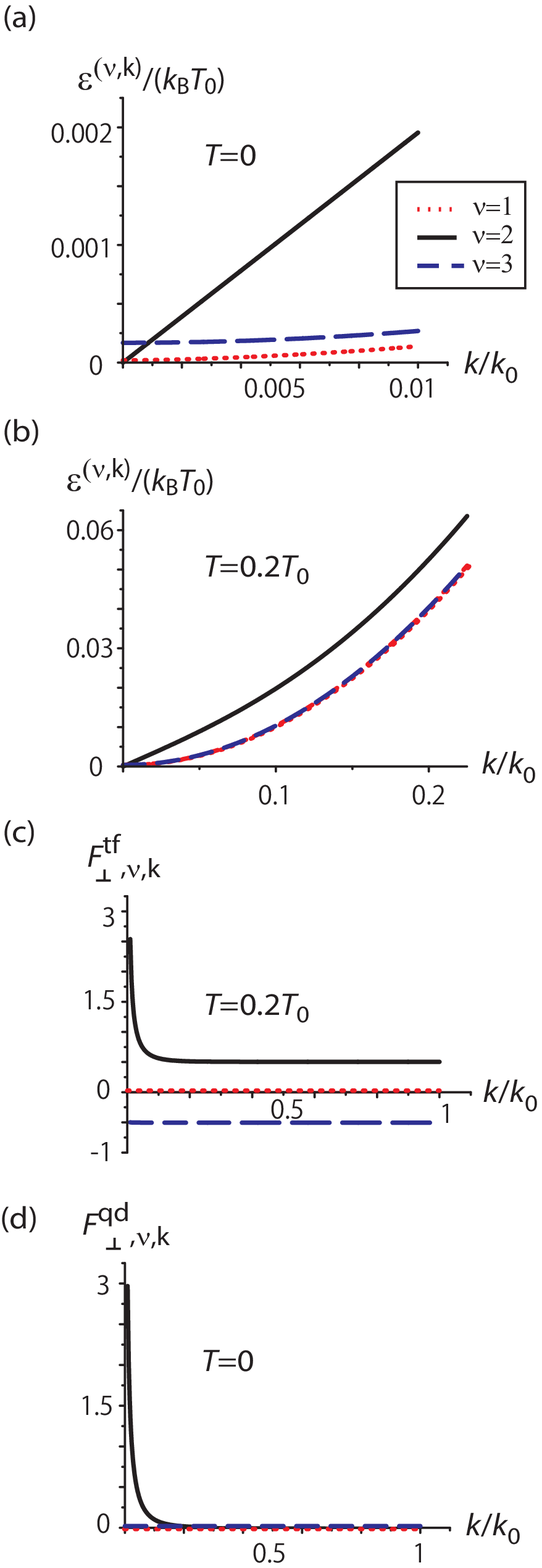}
  \caption{(Color online) Energy spectra of excitation modes at (a) $T=0$ and (b) $T=0.2~T_0$, and the contributions of these modes to the transverse magnetization $F^\mathrm{nc}_\perp$ of the noncondensate due to (c) thermal fluctuations $F^{\rm tf}_{\perp,\nu,\K}$ ($T=0.2~T_0$) and (d) quantum depletion $F^{\rm qd}_{\perp,\nu,\K}$ ($T=0$) [see Eqs.~(\ref{eq:Fthermal}) and (\ref{eq:Ftf,Fqd})] for $q=|c_1|n$ and $n=1\times10^{12}~\mathrm{cm}^{-3}$. The magnitude of wavevector $k=|\K|$ is measured in units of $k_0\equiv \sqrt{2Mk_{\rm B}T_0}/\hbar$. There are in total three excitation modes labeled by $\nu=1, 2$, and 3, which are shown by the dotted, solid, and dashed curves. The energy spectra of $\nu=1$ and $\nu=3$ modes almost coincide in the high-momentum region as shown in (b). The negative values of $F^{\rm tf}_{\perp,\nu=3,\K}$ in (c) imply that the transverse magnetization of this mode is antiparallel to that of the condensate. Note that $F^{\rm tf}_{\perp,\nu=1,\K}$ in (c) and $F^{\rm qd}_{\perp,\nu=1,3,\K}$ in (d) vanish.}
  \label{fig:energy spectrum}
\end{figure}

Figures~\ref{fig:energy spectrum}(a) and \ref{fig:energy spectrum}(b) show the energy spectra $\epsilon^{(\nu,\K)}$ of the three modes ($\nu=1, 2, 3$) at $T=0$ and $0.2~T_0$, respectively. Figures~\ref{fig:energy spectrum}(c) and \ref{fig:energy spectrum}(d) plot $F^{\rm tf}_{\perp,\nu,\K}=\sigma |F^{\rm tf}_{+,\nu,\K}|$ and $F^{\rm qd}_{\perp,\nu,\K}=\sigma |F^{\rm qd}_{+,\nu,\K}|$, respectively, where $\sigma=1$ ($\sigma=-1$) if the magnetization of the condensate and that of the noncondensate are parallel (antiparallel) to each other. It can be seen from Fig.~\ref{fig:energy spectrum}(c)  that the $\nu=1$ mode has no magnetization (dotted line), while the other two have magnetizations parallel ($\nu=2$, solid curve) and antiparallel ($\nu=3$, dashed curve) to that of the condensate. Note here that the excitation energy of the $\nu=2$ mode is higher than that of the $\nu=3$ mode at high momenta [Figs. \ref{fig:energy spectrum}(a) and \ref{fig:energy spectrum}(b)]. Consequently, at high temperatures, the number of thermally excited quasiparticles in the $\nu=2$ mode is smaller than that in the $\nu=3$ mode, leading to the negative $F^\mathrm{nc}_\perp$, which implies that the noncondensate is magnetized in the direction antiparallel to that of the condensate. The above difference between the energy of the $\nu=2$ and $\nu=3$ modes can be explained as follows. If the noncondensed atoms have spin configurations differing from that of the condensate ($\nu=3$), they interact with the condensate only via the direct (Hartree) term. In contrast, if they have the same spin configuration as the condensate ($\nu=2$), both the direct (Hartree) and exchange (Fock) exist, making the excitation energy of the $\nu=2$ mode higher than that of the $\nu=3$ mode.

On the other hand, in the low-momentum regime, the $\nu=2$ mode has a gapless linear dispersion relation, which results in a nonzero $F^{\rm qd}_{\perp,2,\K}$ [Fig.~\ref{fig:energy spectrum}(d)], whereas the $\nu=3$ mode has an energy gap, which suppresses the quantum depletion at absolute zero. Consequently, the magnetization of the noncondensate becomes parallel to that of the condensate in this very low-temperature region. The temperature at which the magnetization of the noncondensate changes its direction is $T_\mathrm{qd}\simeq 0.01~T_0$. Below it, the effect of quantum depletion becomes significant. This crossover temperature is, however, much higher than the energy gap of the $\nu=3$ mode ($\sim |c_1|n/k_\mathrm{B}\simeq9\times 10^{-5}~T_0$), which is a spinor manifestation of the Bose enhancement in the presence of a magnetized condensate.

%%%%%%%%%%%%%%%%%%%%%%%%%%
\section{Effects of Noncondensed Atoms on the Ground State at Absolute Zero}
\label{sec:Effects of noncondensed Atoms on the Ground State at Absolute Zero}
In dilute, weakly interacting Bose gases, the fraction of noncondensed atoms due to quantum depletion at absolute zero is very small. For example, for a uniform scalar BEC of $\Rb$ atoms with atomic density $n=10^{12}~\mathrm{cm}^{-3}$, the quantum depletion at absolute zero is evaluated to be $n^\mathrm{nc}/n=8(na^3)^{1/2}/3\sqrt{\pi}\simeq 5\times 10^{-4}$ (see~\cite{Pethick08}, p.~233) and its effect on the condensate is only to shift the chemical potential. Even for a trapped system, such a small fraction of noncondensed atoms hardly affects the shape of the condensate. For a spinor BEC, however, the quantum depletion significantly alters the magnetism of the condensate as we have discussed at the end of Sec.~\ref{subsec:Finite-temperature Phase Diagram}: the phase boundary between the BA and polar phases shifts due to quantum depletion.

The reason why a minute quantum depletion leads to a significant change in the ground-state magnetism can be understood from the generalized GP equation~(\ref{eq: generalized GP eq. for spinor BEC in Popov approximation}). In the mean-field approximation, the order parameter of a uniform system is determined by the competition between the quadratic Zeeman energy $\sim qi^2\delta_{ij}$ and the spin-dependent interaction $\sim c_1 F^\mathrm{c}_\alpha (f_\alpha)_{ij}$. In the first-order self-consistent approximation, since $|c_1|\ll c_0$, the noncondensed atoms affect the ground-state wave function mainly via the terms $c_0 n^\mathrm{nc}$ and $c_0\tilde{n}_{ij}^*$ in the first line of Eq.~(\ref{eq: generalized GP eq. for spinor BEC in Popov approximation}). Here, the term $c_0 n^\mathrm{nc}$ merely shifts the chemical potential as in the case of a scalar BEC, whereas the term $c_0 \tilde{n}_{ij}^*$ mixes the spinor components $\psi_j$'s, thereby changing the magnetism of the condensate. Note that for the $\Rb$ atoms in $F=1$ hyperfine manifold, the spin-independent interaction is about $200$ times larger than the spin-dependent interaction. Due to such a large ratio $c_0/|c_1|$, the term $c_0 \tilde{n}_{ij}^*$ can have a magnitude comparable to that of the spin-dependent interaction $c_1 F^\mathrm{c}_\alpha(f_\alpha)_{ij}$ between condensed atoms even at absolute zero.

We therefore need to investigate the effect of the anomalous average, which is neglected in the first-order self-consistent approximation, on the ground-state magnetism. To take into account the effects of both the anomalous average and the noncondensate density, we use the perturbative expansion in powers of $\chi^{1/2}$, where $\chi\equiv n^\mathrm{nc}/n$ is the noncondensate fraction which is small at absolute zero. This approach was first proposed by Castin and Dum for scalar BECs~\cite{Castin98}, and we generalize it to spinor gases.

%%%%%%%%%%%%%%%%%%%%%%%%
\subsection{$\chi^{1/2}$ perturbative expansion}
\label{subsec:Perturbative Expansion}
According to Penrose and Onsager~\cite{Penrose56}, the condensate wave function, which plays the role of the order parameter, is defined as the eigenfunction of the one-particle reduced density matrix $\rho^1_{ij}(\R,\R',t)\equiv \langle \creat{j}(\R') \annih{i}(\R) \rangle_t$ with a macroscopic eigenvalue:
\begin{equation}
\int \, \text{d}\R'\,\sum_j \rho^1_{ij}(\R,\R',t)\varphi_j(\R',t)=N^\mathrm{c}(t)\varphi_i(\R,t),
\end{equation}
where $N^\mathrm{c}(t)$ is the number of particles in the condensate, and $\varphi_i(\R,t)\; (i=1,0,-1)$ is the condensate wave function, which is normalized as
\begin{equation}
\integral \sum\limits_i |\varphi_i(\R,t)|^2=1.
\end{equation}
The condensate wave function is conventionally defined as $\phi_i(\R,t)=\sqrt{N^\mathrm{c}(t)}\varphi_i(\R,t)$. However, throughout Sec.~\ref{sec:Effects of noncondensed Atoms on the Ground State at Absolute Zero}, the term \textquotedblleft condensate wave function" refers to $\varphi_i(\R,t)$.

The field operator is then separated into the condensate and noncondensate parts:
\begin{equation}
\annih{i}(\R)=\varphi_i(\R,t)\hat{a}(t)+\ann{i}(\R,t).
\label{eq: field operator decomposition}
\end{equation}
The noncondensate fraction is defined as
\begin{equation}
\chi(t)\equiv \frac{N^\mathrm{nc}(t)}{N} \simeq \frac{N^\mathrm{nc}(t)}{N^\mathrm{c}(t)}=\frac{\integral \sum\limits_{i} \langle\cre{i}(\R,t)\ann{i}(\R,t) \rangle}{\langle \hat{a}^\dagger(t) \hat{a}(t)\rangle},
\end{equation}
and, thus, we have
\begin{equation}
\frac{\ann{i}(\R,t)}{\varphi_i(\R,t)\hat{a}(t)}\sim \sqrt{\frac{n^\mathrm{nc}(t)}{n^\mathrm{c}(t)}}=\sqrt{\chi(t)}.
\label{eq:delta/phi}
\end{equation}

We start with the Heisenberg equation of motion for the operator $\hat{a}^\dagger(t)\ann{i}(\R,t)$:
\begin{align}
i\hbar \frac{d}{dt}\left(\hat{a}^\dagger(t)\ann{i}(\R,t)\right)=\ &i\hbar\frac{\partial}{\partial t}\left(\hat{a}^\dagger(t)\ann{i}(\R,t)\right)\nonumber\\
&+\left[\hat{a}^\dagger(t)\ann{i}(\R,t),\Hamil{}\right],
\label{eq: Heisenberg eq for spinor BEC in Castin formalism}
\end{align}
where $\Hamil{}$ is given by Eq.~(\ref{eq: Spin-1 spinor BEC's Hamiltonian}). 

Expanding the right-hand side of Eq.~(\ref{eq: Heisenberg eq for spinor BEC in Castin formalism}) in powers of $\ann{i}(\R,t)$, and collecting terms of the same order of magnitude, we obtain
\begin{equation}
i\hbar \frac{d}{dt}(\hat{a}^\dagger(t)\ann{i}(\R,t))= \integrals \sum_{n=0}^{4} R_n(\R,\mathbf{s},t),
\label{eq: R_i (i=0,...,4)}
\end{equation}
where $R_n\;(n=0,\dots,4)$ is the sum of terms that contain the $n$-th power of $\ann{i}(\R,t)$. The explicit expressions for $R_n$ ($n=0, 1$, and 2) are given below. (The terms $R_3$ and $R_4$ are irrelevant when one makes an expansion up to the order of $\chi^1$.)

The first term on the right-hand side of Eq.~(\ref{eq: Heisenberg eq for spinor BEC in Castin formalism}) can be written in terms of the field operator $\annih{i}(\R,t)$ by using the following expressions:
\begin{align}
\roundt\hat{a}^\dagger(t)=& \integrals\sum_j \left[\roundt \varphi_j(\s,t)\right]\creat{j}(\s,t), \\
\roundt\ann{i}(\R,t)=& \integrals \sum_j\left[\roundt Q_{ij}(\R,\s,t)\right]\annih{j}(\s,t).
\end{align}
In the following, the argument $t$ of $\annih{i}(\R,t),\ann{i}(\R,t),\hat{a}(t)$ is occasionally omitted to save space in lengthy expressions.

By using Eq.~(\ref{eq: Spin-1 spinor BEC's Hamiltonian}), the last term in Eq.~(\ref{eq: Heisenberg eq for spinor BEC in Castin formalism}) is rewritten as
\begin{align}
&\Bigg[\hat{a}^\dagger(t)\ann{i}(\R,t),\integrals \Bigg\{\sum_{j,l}\creat{j}(\s)(h_0)_{jl}\annih{l}(\s)\nonumber\\
&+\frac{c_0}{2}\sum_{j,l}\creat{j}(\s)\creat{l}(\s)\annih{l}(\s)\annih{j}(\s) \nonumber\\
&+\frac{c_1}{2}\sum_{\alpha,j,k,g,l}(f_\alpha)_{jk}(f_\alpha)_{gl}\creat{j}(\s)\creat{g}(\s)\annih{l}(\s)\annih{k}(\s)\Bigg\} \Bigg].
\label{eq:*2}
\end{align}
This commutator can be calculated by using the relations
\begin{align}
[\hat{a}^\dagger(t),\annih{i}(\s,t)]=&-\varphi_i(\s,t),\\
[\ann{i}(\R,t),\creat{j}(\s,t)]=&\ Q_{ij}(\R,\s,t),
\end{align}
which follow from the orthogonality between the condensate and noncondensate.

Substituting all the above results into Eq.~(\ref{eq: Heisenberg eq for spinor BEC in Castin formalism}), the Heisenberg equation of motion can be rewritten as
\begin{widetext}
\begin{align}
i\hbar\frac{d}{dt}(\hat{a}^\dagger(t)\ann{i}(\R,t))=&\integrals \Bigg\{\sum_j \Bigg[i\hbar\roundt\varphi_j(\s,t)\creat{j}(\s)\ann{i}(\R)+i\hbar\roundt Q_{ij}(\R,\s,t)\hat{a}^\dagger\annih{j}(\s) \Bigg]\nonumber\\
&+\sum_{j,l}\Bigg[-(h_0)_{jl}\varphi_l(\s)\creat{j}(\s)\ann{i}(\R)+Q_{ij}(\R,\s)\hat{a}^\dagger(h_0)_{jl}\annih{l}(\s)\nonumber\\
&+c_0\Big(-\varphi_l(\s)\creat{j}(\s)\creat{l}(\s)\annih{j}(\s)\ann{i}(\R)+Q_{ij}(\R,\s)\hat{a}^\dagger \creat{l}(\s)\annih{l}(\s)\annih{j}(\s)\Big)\Bigg] \nonumber\\
&+c_1\sum_{\alpha,j,k,g,l}(f_\alpha)_{jk}(f_\alpha)_{gl}\Bigg[-\varphi_{l}(\s)\creat{j}(\s)\creat{g}(\s)\annih{k}(\s)\ann{i}(\R)+Q_{ij}(\R,\s)\hat{a}^\dagger\creat{g}(\s)\annih{l}(\s)\annih{k}(\s)\Bigg] \Bigg\}.
\label{eq: calculating LHS of Heisenberg equation}
\end{align}
%\end{widetext}
Next, by substituting Eq.~\eqref{eq: field operator decomposition} into Eq.~\eqref{eq: calculating LHS of Heisenberg equation} and collecting terms according to the power of the noncondensate operator $\ann{i}$, we obtain $R_n$'s $(n=0, 1, 2)$ in Eq.~(\ref{eq: R_i (i=0,...,4)}) as follows. 
%\begin{widetext}
%\begin{subequations}
\begin{align}
R_0(\R,\s,t)=&\,\adagg\ahat \sum_j Q_{ij}(\R,\s)\Bigg\{\sum_l\Bigg[-i\hbar\delta_{jl}\roundt+(\hat{h}_0)_{jl}+c_0(\adagg\ahat-1)\delta_{jl}\sum_k|\varphi_k(\s)|^2\Bigg]\varphi_l(\s,t) \nonumber\\
&+c_1\sum_{\alpha,g,k,l}(f_\alpha)_{jk}(f_\alpha)_{gl}(\adagg\ahat-1)\varphi_g^*(\s)\varphi_{l}(\s)\varphi_{k}(\s)\Bigg\},
\label{eq:expression for R0}
\end{align}
\begin{align}
R_1(\R,\s,t)=&\sum_j \Bigg\{i\hbar\roundt\varphi_j(\s,t)\varphi_j^*(\s)\adagg\ann{i}(\R)+i\hbar\roundt Q_{ij}(\R,\s,t)\adagg\ann{j}(\s)\Bigg\}\nonumber\\
&+\sum_{j,l}\Bigg\{-(h_0)_{jl}\varphi_l(\s)\varphi_j^*(\s)\adagg\ann{i}(\R)+Q_{ij}(\R,\s)\adagg(h_0)_{jl}\ann{l}(\s)\nonumber\\
&+c_0\Big[-\varphi_j^*(\s)\varphi_l^*(\s)\varphi_l(\s)\varphi_j(\s)\adagg\adagg\ahat\ann{i}(\R)+Q_{ij}(\R,\s)\varphi_l^*(\s)\varphi_l(\s)\adagg\adagg\ahat\ann{j}(\s)\nonumber\\
&+Q_{ij}(\R,\s)\varphi_l(\s)\varphi_j(\s)\cre{l}(\s)\adagg\ahat\ahat+Q_{ij}(\R,\s)\varphi_l^*(\s)\varphi_j(\s)\adagg\adagg\ahat\ann{l}(\s)\Big] \Bigg\}\nonumber\\
&+c_1\sum_{\alpha,j,k,g,l}(f_\alpha)_{jk}(f_\alpha)_{gl}\Bigg\{-\varphi_j^*(\s)\varphi_g^*(\s)\varphi_{l}(\s)\varphi_{k}(\s)\adagg\adagg\ahat\ann{i}(\R)+Q_{ij}(\R,\s)\varphi_{l}(\s)\varphi_{k}(\s)\cre{g}(\s)\adagg\ahat\ahat\nonumber\\
&+Q_{ij}(\R,\s)\varphi_{g}^*(\s)\varphi_{k}(\s)\adagg\adagg\ahat\ann{l}(\s)+Q_{ij}(\R,\s)\varphi_{g}^*(\s)\varphi_{l}(\s)\adagg\adagg\ahat\ann{k}(\s)\Bigg\},
\end{align}
\begin{align}
R_2(\R,\s,t)=&\sum_j i\hbar\roundt \varphi_j(\s,t)\cre{j}(\s)\ann{i}(\R)\nonumber\\
&+\sum_{j,l}\Bigg\{-(h_0)_{jl}\varphi_l(\s)\cre{j}(\s)\ann{i}(\R)+c_0\Bigg[-2\varphi_l^*(\s)\varphi_l(\s)\varphi_j(\s)\cre{j}(\s)\adagg\ahat\ann{i}(\R)-\varphi_j^*(\s)\varphi_l^*(\s)\varphi_l(\s)\adagg\adagg\ann{j}(\s)\ann{i}(\R) \nonumber\\
&+Q_{ij}(\R,\s)\Big(\varphi_j(\s)\cre{l}(\s)\adagg\ahat\ann{l}(\s)+\varphi_l^*(\s)\adagg\adagg\ann{l}(\s)\ann{j}(\s)+\varphi_l(\s)\cre{l}(\s)\adagg\ahat\ann{j}(\s)\Big)\Bigg]\Bigg\} \nonumber\\
&+c_1\sum_{\alpha,j,k,g,l}(f_\alpha)_{jk}(f_\alpha)_{gl}\Bigg\{-2\varphi_g^*(\s)\varphi_{l}(\s)\varphi_{k}(\s)\cre{j}(\s)\adagg\ahat\ann{i}(\R)-\varphi_g^*(\s)\varphi_{j}^*(\s)\varphi_{l}(\s)\adagg\adagg\ann{k}(\s)\ann{i}(\R) \nonumber\\
&+Q_{ij}(\R,\s)\Big[\varphi_g^*(\s)\adagg\adagg\ann{l}(\s)\ann{k}(\s)+\varphi_{l}(\s)\cre{g}(\s)\adagg\ahat\ann{k}(\s)+\varphi_{k}(\s)\cre{g}(\s)\adagg\ahat\ann{l}(\s)\Big]\Bigg\}.
\end{align}
%\end{subequations}
\end{widetext}

To make a systematic expansion in powers of $\sqrt{\chi}$, we expand the condensate wave function $\varphi_i(\R,t)$ and the generalized noncondensate operator, defined by $\hat{\Lambda}_i(\R,t)\equiv \hat{a}^\dagger(t)\ann{i}(\R,t)/\sqrt{N^\mathrm{c}}$, as follows:
\begin{equation}
\varphi_i(\R,t)=\overbrace{\underbrace{\varphi_i^{(0)}(\R,t)+\sqrt{\chi}\,\delta\varphi_i^{(1)}(\R,t)}_{\varphi_i^{(1)}(\R,t)}+\chi\,\delta\varphi_i^{(2)}(\R,t)}^{\varphi_i^{(2)}(\R,t)}+\dots,
\label{eq: expansion of phi}
\end{equation}
\begin{equation}
\hat{\Lambda}_i(\R,t)=\overbrace{\underbrace{\hat{\Lambda}_i^{(0)}(\R,t)+\sqrt{\chi}\,\delta\hat{\Lambda}_i^{(1)}(\R,t)}_{\hat{\Lambda}_i^{(1)}(\R,t)}+\chi\,\delta\hat{\Lambda}_i^{(2)}(\R,t)}^{\hat{\Lambda}_i^{(2)}(\R,t)}+\dots,
\label{eq: expansion of lambda}
\end{equation}
where we define
\begin{align}
\varphi_i^{(0)}(\R,t)\equiv&\lim\limits_{\chi \to 0} \varphi_i(\R,t), \\
\delta\varphi_i^{(1)}(\R,t)\equiv&\lim\limits_{\chi \to 0} (\varphi_i(\R,t)-\varphi_i^{(0)}(\R,t))/\sqrt{\chi},\\
\varphi_i^{(1)} \equiv&\ \varphi_i^{(0)} + \sqrt{\chi} \delta\varphi_i^{(1)},
\label{eq:expansion}
\end{align}
and so on. Note that the perturbative expansions in Eqs.~\eqref{eq: expansion of phi} and \eqref{eq: expansion of lambda} hold only if the system does not undergo a quantum phase transition as the total number density $n$ ($\propto \chi^2$) is increased from zero to the final value, during which the order parameter and energy spectrum change smoothly with $\sqrt{\chi}$.

By expanding both sides of Eq.~(\ref{eq: Heisenberg eq for spinor BEC in Castin formalism}) up to the order of $\chi^0$, $\chi^{1/2}$ and $\chi^1$ successively, and using the orthogonality relation (see Appendix~\ref{sec: Derivation of Eq. expectation value identity for Heisenberg equation in Castin} for the derivation)
\begin{equation}
\langle \hat{a}^\dagger(t)\ann{i}(\R,t)\rangle =0,
\label{eq: expectation value identity for Heisenberg equation in Castin}
\end{equation}
we obtain the equations that must be satisfied by the condensate wave functions at different orders $\varphi_i^{(0)}(\R,t)$, $\varphi_i^{(1)}(\R,t)$, $\varphi_i^{(2)}(\R,t)$ and the lowest-order noncondensate operator $\Lambda_i^{(0)}(\R,t)$. Here, we outline the procedures 1-3 for deriving those equations.

1. Expansion up to the order of $\chi^0$. \\
By using Eq.~(\ref{eq: expectation value identity for Heisenberg equation in Castin}) in the left-hand side of Eq.~(\ref{eq: R_i (i=0,...,4)}) and neglecting the terms that contains $\ann{i}(\R,t)$, we obtain $\langle \integrals R_0(\R,\s,t) \rangle=0$, which leads to the time-dependent GP equation for $\varphi_i^{(0)}(\R,t)$ (see Appendix~\ref{sec: derivation of eq: order chi 0} for the derivation):
\begin{align}
&\sum_{j}\left[ -i\hbar\delta_{ij}\frac{\partial}{\partial t}+(\hat{h}_0)_{ij}+c_0 N\delta_{ij} \sum_k|\varphi_k^{(0)}|^2 \right] \varphi_j^{(0)}(\R,t) \nonumber \\
&+c_1 N \sum_{\alpha,j,k,l}(f_\alpha)_{ij}(f_\alpha)_{kl}\varphi_k^{(0)*}\varphi_{l}^{(0)}\varphi_{j}^{(0)}\nonumber\\
&=\eta^{(0)}(t)\varphi_i^{(0)}(\R,t).
\label{eq: order chi 0}
\end{align}
Here $\eta^{(0)}(t)$ is an arbitrary real function relating $\varphi_i^{(0)}$ to $\varphi_i^{(0)'}$ through a unitary transformation $\varphi_i^{(0)'}(\R,t)=\varphi_i^{(0)}(\R,t)\exp\Big[i\int\limits_{0}^{t}\mathrm{d}t'\eta^{(0)}(t')/\hbar\Big]$. The dynamics of $\varphi_i^{(0)'}(\R,t)$ is governed by the equation that is similar to Eq.~\eqref{eq: order chi 0} but with the right-hand side being replaced by 0.

2. Expansion up to the order of $\chi^{1/2}$. \\
Similarly, we can obtain the equation for the condensate wave function at the next order, $\varphi_i^{(1)}(\R,t)$. It turns out that the equation that must be satisfied by $\varphi_i^{(1)}(\R,t)$ is the same as that for $\varphi_i^{(0)}(\R,t)$, i.e., Eq.~\eqref{eq: order chi 0}. In other words, the condensate wave function does not change at this order. Also, at this order, the equation of motion for the noncondensate operator at the lowest order, $\hat{\Lambda}_i^{(0)}(\R,t)$, can be obtained by expanding both sides of Eq.~(\ref{eq: R_i (i=0,...,4)}) up to the order of $\chi^{1/2}$. It is the time-dependent BdG equation (see Appendix~\ref{sec: derivation of eq: order chi 1/2} for the derivation):
\begin{equation}
i\hbar\frac{d}{dt}\hat{\Lambda}_i^{(0)}(\R,t)=\sum_j\Big[A_{ij} \hat{\Lambda}_j^{(0)}(\R,t) + B_{ij} \hat{\Lambda}_j^{(0)\dagger}(\R,t)\Big],
\label{eq: order chi 1/2}
\end{equation}
where
\begin{align}
A_{ij}=&(\hat{h}_0)_{ij}+\left[-\eta^{(0)}+c_0 N \sum_k |\varphi_k^{(0)}|^2\right]\delta_{ij} \nonumber\\
&+ \sum_{k,l}\Bigg\{c_1 N (f_\alpha)_{ij}(f_\alpha)_{kl}\varphi_k^{(0)*}\varphi_{l}^{(0)}\nonumber\\
&+ \hat{Q}^{(0)}_{ik}\circ \Bigg[ c_0 N \varphi_k^{(0)}\varphi_l^{(0)*} \nonumber\\
&+c_1 N \sum_{h,g}(f_\alpha)_{kh}(f_\alpha)_{gl}\varphi_{g}^{(0)*}\varphi_h^{(0)} \Bigg]\circ \hat{Q}^{(0)}_{lj}\Bigg\},\\
B_{ij}=&\sum_{k,l}\Bigg\{\hat{Q}^{(0)}_{ik}\circ\Bigg[ c_0 N \varphi_k^{(0)}\varphi_l^{(0)}\nonumber\\
& +c_1 N \sum_{h,g}(f_\alpha)_{kh}(f_\alpha)_{lg}\varphi_{h}^{(0)}\varphi_{g}^{(0)} \Bigg] \circ \hat{Q}^{(0)*}_{lj}\Bigg\}.
\label{eq: BdG matrices in Castin}
\end{align}
Here, $\hat{Q}^{(0)}_{ij}$ is the projection operator onto the subspace orthogonal to the condensate wave function at the lowest order, $\varphi_i^{(0)}(\R)$. Its action on an arbitrary vector component $f_j(\R)$ is given by
\begin{align}
\hat{Q}^{(0)}_{ij}\circ f_j(\R)=\sum_j \integral' Q^{(0)}_{ij}(\R,\R') f_j(\R'),
\end{align}
where $Q^{(0)}_{ij}(\R,\R')=\delta_{ij}\delta(\R-\R')-\varphi^{(0)}_i(\R)\varphi^{(0)*}_j(\R')$. For uniform systems, however, the elementary excitations are plane waves with nonzero momenta, which are orthogonal to the condensate wave function, and, therefore, the projection operator $\hat{Q}_{ij}^{(0)}$ can be omitted.

3. Expansion up to the order of $\chi^1$.\\
By using Eq.~(\ref{eq: expectation value identity for Heisenberg equation in Castin}) and keeping the terms on the right-hand side of Eq.~(\ref{eq: R_i (i=0,...,4)}) up to the order of $\chi^1$, we obtain the generalized GP equation for the condensate wave function at this order $\varphi_i^{(2)}(\R)$, in which the effects of both the noncondensate number density and the anomalous average are included (see Appendix~\ref{sec: derivation of eq: order chi 1} for the derivation):
\begin{align}
&\sum_j\Bigg\{\Bigg[ -i\hbar\frac{\partial}{\partial t}\delta_{ij}+(\hat{h}_0)_{ij}+c_0 N^\mathrm{c}\delta_{ij} \sum_k|\varphi_k^{(2)}|^2 \Bigg] \varphi_j^{(2)}(\R,t) \nonumber \\
&+c_0 \left[ \tilde{n}_{jj}\varphi_i^{(2)} + \tilde{n}_{ij}^*\varphi_j^{(2)} +\tilde{m}_{ij}^{(R)}\varphi_j^{(2)*} \right] \Bigg\}\nonumber \\
&+c_1\sum_{\alpha,j,k,l}(f_\alpha)_{ij}(f_\alpha)_{kl}\Bigg\{ N^\mathrm{c} \varphi_k^{(2)*}\varphi_l^{(2)}\varphi_{j}^{(2)} \nonumber \\
&+\left[ \tilde{n}_{kl}\varphi_j^{(2)} +\tilde{n}_{jk}^*\varphi_l^{(2)} + \tilde{m}_{jl}^{(R)}\varphi_k^{(2)*} \right]\Bigg\}-\mathcal{F}_i(\R)\nonumber\\
&= \eta^{(2)}(t)\varphi_i^{(2)}(\R,t).
\label{eq: order chi 1}
\end{align}
Here, $N^\mathrm{c}<N$ is the number of atoms in the condensate, $\tilde{m}_{ij}\order{R}$ is the renormalized anomalous average, and $\mathcal{F}_i(\R)$ is defined as
\begin{align}
\mathcal{F}_i(\R)\equiv& \int \, \text{d}\s \Bigg\{ c_0 N^\mathrm{c} \left(\sum_k|\varphi\order{0}_k(\s)|^2 \right)\nonumber\\
&\sum_j\Big[ \tilde{n}_{ij}^*(\R,\s)\varphi_j\order{0}(\s)+ \tilde{m}_{ij}(\R,\s)\varphi^{(0)*}_j(\s) \Big]\nonumber\\
&+c_1 N^\mathrm{c}\sum_{\alpha,j,k,g,l} (f_\alpha)_{jk}(f_\alpha)_{gl}\varphi^{(0)*}_g(\s)\varphi\order{0}_{l}(\s) \nonumber\\
&\Big[\tilde{n}_{ij}^*(\R,\s)\varphi\order{0}_{k}(\s) + \tilde{m}_{ik}(\R,\s)\varphi^{(0)*}_j(\s)\Big] \Bigg\}.
\label{eq: f_m(R)}
\end{align}
The matrix elements of the noncondensate number density $\tilde{n}_{ij}(\R,\s,t)$ and the anomalous average $\tilde{m}_{ij}(\R,\s,t)$ are defined in terms of the noncondensate operator at the lowest order $\hat{\Lambda}\order{0}_i(\R,t)$ as
\begin{align}
\tilde{n}_{ij}(\R,\s,t)\equiv&\ \langle \hat{\Lambda}^{(0)\dagger}_i(\R,t) \hat{\Lambda}\order{0}_j(\s,t) \rangle, \label{eq:n-tilde} \\
\tilde{m}_{ij}(\R,\s,t)\equiv&\ \langle \hat{\Lambda}\order{0}_i(\R,t) \hat{\Lambda}\order{0}_j(\s,t) \rangle.
\label{eq:m-tilde}
\end{align}

From the time-dependent generalized GP equation for the condensate wave function $\varphi_i\order{2}(\R,t)$ [Eq.~(\ref{eq: order chi 1})] and the time-dependent BdG equation for the noncondensate operator $\hat{\Lambda}\order{0}_i(\R,t)$ [Eq.~(\ref{eq: order chi 1/2})], both the dynamics and thermal equilibrium properties of a spinor Bose gas can be obtained.

Using the Bogoliubov transformation, the time evolution of the noncondensate operator $\hat{\Lambda}\order{0}_i(\R,t)$ can be expressed as
\begin{equation}
\begin{pmatrix}
\hat{\Lambda}\order{0}_i(\R,t)\\
\hat{\Lambda}^{(0)\dagger}_i(\R,t)\\
\end{pmatrix}
= \sum_{\lambda} \hat{b}_{\lambda}
\begin{pmatrix}
u_i^{(\lambda)}(\R,t)\\
v_i^{(\lambda)}(\R,t)\\
\end{pmatrix}
+ \hat{b}_{\lambda}^\dagger
\begin{pmatrix}
v_i^{(\lambda)*}(\R,t)\\
u_i^{(\lambda)*}(\R,t)\\
\end{pmatrix},
\end{equation}
where $\hat{b}_{\lambda}$ and $\hat{b}_{\lambda}^\dagger$ are the creation and annihilation operators of the excitation mode labeled by index $\lambda$, and the coefficients $u_i^{(\lambda)}(\R,t)$ and $v_i^{(\lambda)}(\R,t)$ are given by
\begin{equation}
\begin{pmatrix}
u_i^{(\lambda)}(\R,t)\\
v_i^{(\lambda)}(\R,t)\\
\end{pmatrix}
=e^{-i\epsilon\order{\lambda} t/\hbar}
\begin{pmatrix}
u_i^{(\lambda)}(\R)\\
v_i^{(\lambda)}(\R)\\
\end{pmatrix}.
\end{equation}
Here, $u_i\order{\lambda}(\R)$ and $v_i\order{\lambda}(\R)$ are the solutions of Eq.~(\ref{eq: BdG equation for spinor BEC}) with $A_{ij}$ and $B_{ij}$ replaced by those in Eq.~(\ref{eq: BdG matrices in Castin}), and $\epsilon\order{\lambda}$ is the energy of the excitation mode $\lambda$.

%%%%%%%%%%%%%%%%%%%%%%%%%%%%%
\subsection{A uniform Bose gas in thermal equilibrium}
\label{subsec: A uniform Bose gas in thermal equilibrium}
We apply the results in Sec. \ref{subsec:Perturbative Expansion} to a uniform $\Rb$ condensate in thermal equilibrium near absolute zero. In thermal equilibrium, the condensate wave function is time-independent $\varphi\order{2}_i(\R)$, while the occupation numbers of excitation modes are given by the Bose-Einstein distribution
\begin{equation}
\langle \hat{b}_\lambda^\dagger \hat{b}_\lambda \rangle = f(\epsilon\order{\lambda}) \equiv \frac{1}{e^{[\epsilon\order{\lambda}-\mu]/(k_BT)}-1}.
\end{equation}
Here, the chemical potential $\mu$ is taken to be the eigenenergy of the condensate wave function within an error of the order of $1/N$, where $N$ is the total number of particles. The number density and anomalous average of noncondensed atoms are given by
\begin{align}
\tilde{n}_{ij}(\R,\s)=&\sum_{\epsilon^{(\lambda)}>0}\Big\{u_i^{(\lambda)*}(\R) u_j^{(\lambda)}(\s)f(\epsilon^{(\lambda)})\nonumber\\
&+v_i^{(\lambda)}(\R) v_j^{(\lambda)*}(\s)\Big[f(\epsilon^{(\lambda)})+1\Big] \Big\}, \nonumber\\
\tilde{m}_{ij}(\R,\s)=&\sum_{\epsilon^{(\lambda)}>0}\Big\{v_i^{(\lambda)*}(\R) u_j^{(\lambda)}(\s)f(\epsilon^{(\lambda)})\nonumber\\
&+u_i^{(\lambda)}(\R) v_j^{(\lambda)*}(\s)\Big[f(\epsilon^{(\lambda)})+1\Big] \Big\}.
\label{eq:m-tilde 2}
\end{align}

For the uniform system under consideration, the condensate wave function $\varphi_i$ is spatially uniform, while the excitation modes take the form of plane waves:
\begin{subequations}
\label{eq:plane wave 2}
\begin{align}
u_j^{(\lambda)}(\R)=\ &u_j^{(\nu,\K)}e^{i\K.\R}, \\
v_j^{(\lambda)}(\R)=\ &v_j^{(\nu,\K)}e^{i\K.\R},
\end{align}
\end{subequations}
where $\K$ is the wave vector, and $\nu$ is an additional index to distinguish excitation modes. The term $\mathcal{F}_i(\R)$ in Eq~\eqref{eq: f_m(R)} then has the following form:
\begin{align}
\mathcal{F}_i(\R)\propto \frac{1}{\Omega} \sum_{\substack{\K,\nu \\ \epsilon>0}} \int \, \text{d}\R' e^{\pm i\K.\R'} \propto  \sum_{\substack{\K,\nu \\ \epsilon>0}}\delta_{\K,0},
\end{align}
i.e., the nonzero contribution arises only from the excitation modes with zero momentum and positive energy, and it is vanishingly small in the thermodynamic limit.

The set of coupled equations concerning the condensate and excitations in thermal equilibrium is then given as follows:

1. The GP equation for the lowest-order condensate wave function $\varphi_i\order{0}$:
\begin{align}
\left(qi^2+c_0 n\right) \varphi_i^{(0)}+c_1\sum_{\alpha,j}(F_\alpha)(f_\alpha)_{ij}\varphi_{j}^{(0)}=\mu^{(0)}\varphi_i^{(0)},
\label{eq:phi0}
\end{align}
where $\mu^{(0)}$ is the lowest-order chemical potential, $F_\alpha \equiv n\sum\limits_{i,j}\varphi_i^{(0)*}(f_\alpha)_{ij}\varphi_j^{(0)}$ ($\alpha=x,y,z$) are the components of the lowest-order spin density vector, and $\varphi_i^{(0)}$ is normalized to unity:
\begin{equation}
\sum_i |\varphi_i\order{0}|^2=1.
\label{eq:normalize}
\end{equation}

2. The BdG equation for the excitation modes at the lowest order: 
\begin{equation}
\begin{pmatrix}
   A_{ij}(\K) & B_{ij} \\
   -B_{ij}^* & -A_{ij}^*(\K)\\
   \end{pmatrix}
   \begin{pmatrix}
   u^{(\nu,\K)}_j\\
   v^{(\nu,\K)}_j\\
   \end{pmatrix}
   =\epsilon^{(\nu,\K)}
   \begin{pmatrix}
   u^{(\nu,\K)}_i\\
   v^{(\nu,\K)}_i\\
   \end{pmatrix},
\label{eq: BdG Castin}
\end{equation}
where
\begin{align}
A_{ij}(\K)=&\left(\epsilon^0_\K+qi^2+c_0n\right)\delta_{ij}+ c_0 n \varphi_i^{(0)}(\varphi_j^{(0)})^*\nonumber\\
&+ c_1\sum_\alpha\Bigg[(F_\alpha)(f_\alpha)_{ij}\nonumber\\
&+n \sum_{k,l}(f_\alpha)_{il}(f_\alpha)_{kj}(\varphi_{k}^{(0)})^*\varphi_l^{(0)}\Bigg],\\
B_{ij}=&\ c_0 n \varphi_i^{(0)}\varphi_j^{(0)} +\sum_{\alpha,k,l}c_1 n (f_\alpha)_{ik}(f_\alpha)_{jl}\varphi_{k}^{(0)}\varphi_{l}^{(0)}. 
\end{align}
Here, $\epsilon^0_\K=\hbar^2\K^2/(2M)$ is the kinetic energy of a single particle with momentum $\hbar\K$.
 
3. The matrix elements of the noncondensate number density $\tilde{n}_{ij}$ and the anomalous average $\tilde{m}_{ij}$ expressed in terms of the excitation modes:
\begin{align}
\tilde{n}_{ij}=&\sum_\nu \int \frac{\text{d}^3\K}{(2\pi)^3}\,\Big\{u_i^{(\nu,\K)*} u_j^{(\nu,\K)}f^{(0)}(\epsilon^{(\nu,\K)})\nonumber\\
&+v_i^{(\nu,\K)} v_j^{(\nu,\K)*}\left[f^{(0)}(\epsilon^{(\nu,\K)})+1\right] \Big\},\\
\tilde{m}_{ij}=&\sum_\nu \int \frac{\text{d}^3\K}{(2\pi)^3}\,\Big\{v_i^{(\nu,\K)*} u_j^{(\nu,\K)}f^{(0)}(\epsilon^{(\nu,\K)})\nonumber\\
&+u_i^{(\nu,\K)} v_j^{(\nu,\K)*}\left[f^{(0)}(\epsilon^{(\nu,\K)})+1\right] \Big\},
\end{align}
where $f^{(0)}(\epsilon)=1/\{\exp[(\epsilon-\mu^{(0)})/(k_\mathrm{B}T)]-1\}$. 

4. The generalized GP equation for the condensate wave function $\varphi_i\order{2}$ at the order of $\chi^1$ (Note that $\varphi_i\order{1}=\varphi_i\order{0}$ as shown above Eq.~\eqref{eq: order chi 1/2}):
\begin{align}
&[qi^2+c_0(n^\mathrm{c}+n^\mathrm{nc})] \varphi_i^{(2)}+c_0 \sum_j\Big[\tilde{n}_{ij}^*\varphi_j^{(2)} +\tilde{m}_{ij}^{(R)}\varphi_j^{(2)*}\Big] \nonumber \\
&+c_1\sum_{\alpha,j}\Bigg[(F^\mathrm{c}_\alpha+F^\mathrm{nc}_\alpha)(f_\alpha)_{ij}\varphi_{j}^{(2)} \nonumber \\
&+\sum_{k,l}(f_\alpha)_{ij}(f_\alpha)_{kl}\left(\tilde{n}_{kj}\varphi_l^{(2)} + \tilde{m}_{jl}^{(R)}\varphi_k^{(2)*} \right) \Bigg]= \mu^{(2)}\varphi_i^{(2)},
\label{eq: gGP Castin}
\end{align}
where $F^\mathrm{c}_\alpha \equiv n^\mathrm{c}\sum\limits_{i,j}\varphi_i^{(2)*}(f_\alpha)_{ij}\varphi_j^{(2)}$, and $n^\mathrm{nc}$ and $F^\mathrm{nc}_\alpha$ are given by Eqs.~\eqref{eq:nt} and \eqref{eq:nc,nt,Fc,Ft}. Here, $\tilde{m}_{ij}^{(\mathrm{R})}$ is the renormalized anomalous average which is described in Sec. \ref{subsec: Renormalized Anomalous Average Term} below, $\mu^{(2)}$ is the chemical potential at this order, and the order parameter $\varphi_i\order{2}$ is normalized to unity:
\begin{equation}
\sum_i |\varphi_i\order{2}|^2=1.
\end{equation}

5. The number equation for the condensate and noncondensate number densities:
\begin{align}
n=n^\mathrm{c}+n^\mathrm{nc}.
\label{eq: number equation 2}
\end{align}

Note that the generalized GP equation~(\ref{eq: gGP Castin}) for the wave function $\varphi_i^{(2)}$ depends only on the lowest-order noncondensate operator $\Lambda_i^{(0)}(\R)$ via $\tilde{n}_{ij}=\langle\hat{\Lambda}_i^{(0)\dagger}\hat{\Lambda}_j^{(0)}\rangle$ and $\tilde{m}_{ij}=\langle\hat{\Lambda}_i^{(0)}\hat{\Lambda}_j^{(0)}\rangle$ [Eqs.~\eqref{eq:n-tilde}, \eqref{eq:m-tilde}]. This is because the condensate and noncondensate operators are different in the order of magnitude, as shown in Eq.~(\ref{eq:delta/phi}).

%%%%%%%%%%%%%%%%%%%%%%%%%%
\subsection{Renormalized Anomalous Average}
\label{subsec: Renormalized Anomalous Average Term}
The anomalous average term $\tilde{m}_{ij}(\R,\R')$, which is defined in Eq.~(\ref{eq:m-tilde}), represents pair correlation of noncondensed atoms, and can be expressed in terms of the excitation modes as in Eq.~(\ref{eq:m-tilde 2}). However, the summation over all excitation modes in Eq.~(\ref{eq:m-tilde 2}) would give an unphysical divergence. This divergence results from the replacement of the exact interaction by a contact interaction. This replacement amounts to assuming that all short-distance effects of the exact interaction can be encapsulated in one parameter: the $s$-wave scattering length. The effects of all higher-order terms, which represent the multiple-scattering processes involving virtual states with high energies,are implicitly represented by the $s$-wave scattering length. Because $\tilde{m}_{ij}$ is first-order with respect to the interaction, taking into account the effect of pair correlation of noncondensed atoms on the condensate represented by $c_0\tilde{m}_{ij},c_1\tilde{m}_{ij}$, would lead to a double counting of the terms that are second-order with respect to the interaction. This gives rise to the above divergence in the anomalous average term.

To avoid this double counting, we need to go beyond the Born approximation and express the $s$-wave scattering length $a$ up to second-order with respect to the bare interaction. By applying the Lipmann-Schwinger equation (see~\cite{Pethick08}, p.~125) to low-energy collisions between two particles with a contact interaction, we obtain
\begin{equation}
g=\tilde{g}-\frac{\tilde{g}^2}{\Omega}\sum\limits_{k<k_\mathrm{c}}\frac{1}{2\epsilon^0_\K},
\end{equation}
or equivalently,
\begin{equation}
\tilde{g}=g+\frac{g^2}{\Omega}\sum\limits_{k<k_\mathrm{c}}\frac{1}{2\epsilon^0_\K},
\end{equation}
where $g$ is related to the $s$-wave scattering length by $g=4\pi\hbar^2a/M$, while $\tilde{g}$ is the bare contact interaction. Here, $\epsilon^0_\K=\hbar^2\K^2/(2M)$, $\Omega$ is the volume of the system, and $k_\mathrm{c}$ is the cut-off of the momentum.

For spin-1 atoms, there are two $s$-wave scattering lengths $a_0$ and $a_2$ for the total spin $F_\mathrm{total}=0$ and 2 channels, respectively, and therefore, the corresponding coupling constants are given by
\begin{align}
\tilde{g_0}=&g_0+\frac{g_0^2}{\Omega}\sum\limits_{k<k_\mathrm{c}}\frac{1}{2\epsilon^0_\K}, \nonumber\\
\tilde{g_2}=&g_2+\frac{g_2^2}{\Omega}\sum\limits_{k<k_\mathrm{c}}\frac{1}{2\epsilon^0_\K},
\end{align}
where $g_0=4\pi\hbar^2a_0/M$ and $g_2=4\pi\hbar^2a_2/M$.

The spin-independent interaction $\tilde{c}_0$ and spin-dependent interaction $\tilde{c}_1$ are then given by
\begin{align}
\tilde{c}_0=&\frac{\tilde{g}_0+2\tilde{g}_2}{3}, \nonumber\\
\tilde{c}_1=&\frac{\tilde{g}_2-\tilde{g}_0}{3}.
\end{align}

By collecting all second-order terms with respect to the interaction, we obtain
\begin{widetext}
\begin{align}
c_0\sum_j\tilde{m}\order{R}_{ij}\varphi_j^*+c_1\sum_{\alpha,j,k,l}(f_\alpha)_{ij}(f_\alpha)_{kl}\tilde{m}\order{R}_{jl}\varphi_k^*=\ &c_0\sum_j\tilde{m}_{ij}\varphi_j^*+c_1\sum_{\alpha,j,k,l}(f_\alpha)_{ij}(f_\alpha)_{kl}\tilde{m}_{jl}\varphi_k^*\nonumber\\
&+(\tilde{c}_0-c_0)N^\mathrm{c}\left(\sum_j|\varphi_j|^2\right)\varphi_i +(\tilde{c}_1-c_1)N^\mathrm{c}\sum_{\alpha,j,k,l}(f_\alpha)_{ij}(f_\alpha)_{kl}\varphi_k^*\varphi_l\varphi_j \nonumber\\
=&\sum_j\Bigg(c_0\tilde{m}_{ij}+\frac{g_0^2+2g_2^2}{3\Omega}\sum\limits_{k<k_\mathrm{c}}\frac{1}{2\epsilon^0_\K}N\varphi\order{0}_i\varphi\order{0}_j\Bigg)\varphi_j^* \nonumber\\
&+\sum_{\alpha,j,k,l}(f_\alpha)_{ij}(f_\alpha)_{kl}\Bigg(c_1\tilde{m}_{jl}+\frac{g_2^2-g_0^2}{3\Omega}\sum\limits_{k<k_\mathrm{c}}\frac{1}{2\epsilon^0_\K}N\varphi\order{0}_j\varphi\order{0}_l\Bigg)\varphi_k^*.
\label{eq: calculate renormalized anomalous average}
\end{align}
\end{widetext}
Here, in obtaining the last equality, we have replaced $N^\mathrm{c}$ and $\varphi_i$ in the second line of Eq.~\eqref{eq: calculate renormalized anomalous average} by $N$ and $\varphi_i^{(0)}$, respectively. This replacement causes an error of the order of $\chi^{3/2}$, and therefore, is justified up to the order of $\chi^{1}$.

If the mean-field ground state is in the polar phase, i.e., $\bm{\varphi}^{(0)}=(0,1,0)^T$, the matrix element $\tilde{m}\order{R}_{00}$ is given by
\begin{align}
\tilde{m}\order{R}_{00}=&\ \tilde{m}_{00}+c_0n\sum\limits_{\K}\frac{1}{2\epsilon^0_\K}\nonumber\\
=&\ c_0n\sum\limits_{\K}\left[\frac{1}{2\epsilon^0_\K}-\frac{\epsilon^0_\K+c_0n-\epsilon_\K}{(c_0n)^2-(\epsilon^0_\K+c_0n-\epsilon_\K)^2}\right]\nonumber\\
=&\ c_0n\sum_\K\left[\frac{1}{2\epsilon^0_\K}-\frac{1}{2\epsilon_\K}\right],
\label{eq:m00R}
\end{align}
where $\epsilon_\K=\sqrt{\epsilon^0_\K(\epsilon^0_\K+2c_0n)}>\epsilon^0_\K$. Here in the first line of Eq.~\eqref{eq:m00R} we used the fact that $a_2 \simeq a_0$ for $\Rb$. From Eq.~(\ref{eq:m00R}), we find that $\tilde{m}\order{R}_{00}\geq 0$.

%%%%%%%%%%%%%%%%%%%%%%%%%%
\subsection{Ground-state phase diagram}
\label{subsec: Ground-state phase diagram}
\begin{figure}[tbp] % float placement: (h)ere, page (t)op, page (b)ottom, other (p)age
  \centering
  % file name: D:/Paper Drafts (Mar 21st, 2011)/Figures/deviation.EPS
  \includegraphics[bb=0 0 314 221,width=3in,height=2.11in,keepaspectratio]{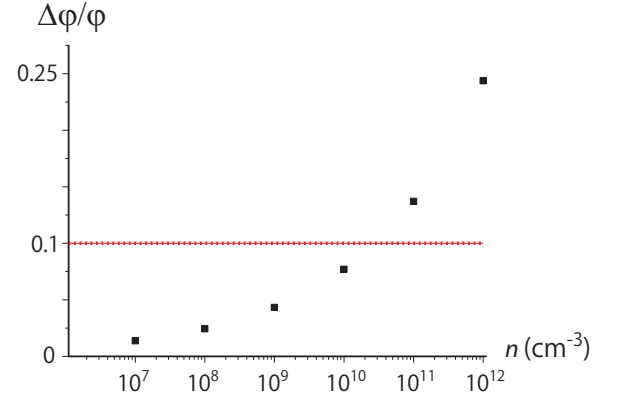}
  \caption{(Color online) Density dependence of the relative shift of the order parameter from the mean-field value: $\Delta\varphi/\varphi\equiv\operatorname*{max}\limits_{i,q}|\varphi_i^{(2)}-\varphi_i^{(0)}|/\operatorname*{max}\limits_{i,q}|\varphi_i^{(0)}|$ (squares). The dotted line $\Delta\varphi/\varphi=0.1$ gives an estimate of the threshold below which the perturbative expansion gives quantitatively reliable results.}
  \label{fig: deviation of phi}
\end{figure}

By numerically solving the set of coupled equations~(\ref{eq:phi0})-(\ref{eq: number equation 2}) we have calculated the ground-state order parameter (i.e., at $T=0$) of a spin-1 ferromagnetic BEC up to the order of $\chi^1$. The parameters of the system are the same as those given in Sec.~\ref{subsec:Finite-temperature Phase Diagram}, namely, those of $\Rb$ atoms. Before discussing the result, let us first evaluate the threshold of the total number density $n_\mathrm{thres}$, beyond which the result obtained by the $\chi^{1/2}$ perturbative expansion deviates so greatly from the mean-field result that the perturbation method no longer gives quantitatively reliable results. Here, we define a measure of the validity of the perturbative expansion as
\begin{align}
\frac{\Delta\varphi}{\varphi} \equiv \frac{\operatorname*{max}\limits_{i,q}|\varphi^{(2)}_i-\varphi^{(0)}_i|}{\operatorname*{max}\limits_{i,q} |\varphi^{(0)}_i|}.
\end{align}
The perturbative expansion is valid if $\Delta\varphi/\varphi \ll 1$. The estimation of the value of $n_\mathrm{thres}$ can be made in the following manner: the large ratio of $c_0/|c_1|\simeq 200$ brings about a significant effect of noncondensed atoms on the spinor condensate; the condition for the effect of noncondensed atoms to be small is therefore given by $\Delta\varphi/\varphi\lesssim 0.1$ or $c_0 n^\mathrm{nc}/(|c_1|n)\lesssim 0.01$ (note that $n\propto |\varphi|^2$); using the expression for the noncondensate fraction $n^\mathrm{nc}/n=8(na^3)^{1/2}/(3\sqrt{\pi})$, we obtain the condition $n\lesssim n_\mathrm{thres}=10^{10}\;{\rm cm}^{-3}$. We have also solved the coupled set of the generalized GP and BdG equations for various values of $n$ to calculate $\Delta \varphi/\varphi$. The result is shown in Fig.~\ref{fig: deviation of phi}, from which we find that the $\chi^{1/2}$ perturbative expansion is valid for $n\lesssim n_\mathrm{thres}=10^{10}\;{\rm cm}^{-3}$.

\begin{figure}[tbp] % float placement: (h)ere, page (t)op, page (b)ottom, other (p)age
  \centering
  % file name: F:/Paper Drafts (May 20th, 2011)/Figures/Fzx_groundstate3.EPS
  \includegraphics[bb=0 0 271 278,width=3in,height=3.08in]{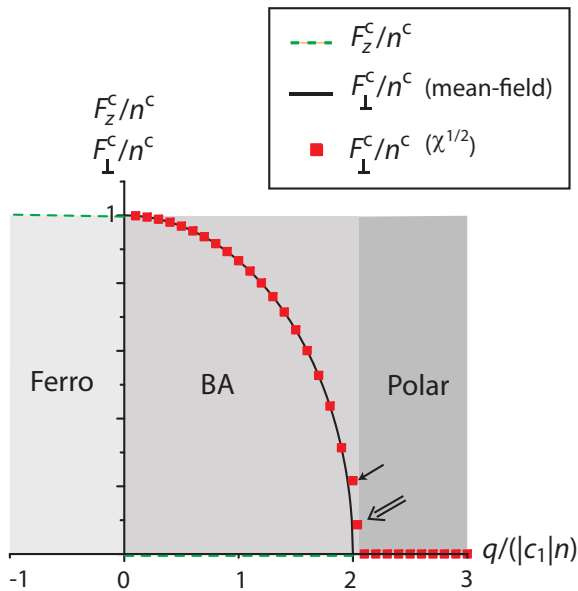}
  \caption{(Color online) $q$ dependence of longitudinal magnetization per condensate particle $F^\mathrm{c}_z/n^\mathrm{c}$ and that of transverse one $F^\mathrm{c}_\perp/n^\mathrm{c}$ for a uniform ferromagnetic BEC at $T=0$. The parameters are those of $\Rb$ with atomic density $n=1\times 10^{10}\ \mathrm{cm}^{-3}$. The squares show the transverse magnetization numerically calculated by using the $\chi^{1/2}$ perturbative expansion, while the solid curve shows the mean-field result given by $F^\mathrm{c}_\perp/n^\mathrm{c}=\sqrt{1-(q/2|c_1|n)^2}\Theta(2|c_1|n-q)\Theta(q)$, where $\Theta(q)$ is the unit-step function. The longitudinal magnetization, which is given by $F^\mathrm{c}_z/n^\mathrm{c}=\Theta(-q)$ and shown by the dashed lines, is the same for the two approximations. The ferromagnetic, BA, and polar phases are shaded in light, medium and dark grey, respectively. The phase boundary between the BA and polar phases lies at $q=q_\mathrm{b}=2.05|c_1|n$. The point indicated by the single arrow shows the value of $F^\mathrm{c}_\perp/n^\mathrm{c}$ at $q/(|c_1|n)=2$. In the mean-field approximation, $F^\mathrm{c}_\perp/n^\mathrm{c}=0$ at $q/(|c_1|n)=2$. The deviation of this point from zero indicates how much the BA phase expands from the mean-field result. The point indicated by the double arrow shows the value of $F^\mathrm{c}_\perp/n^\mathrm{c}$ that lies in the critical parameter region $2|c_1|n<q<q_\mathrm{b}$.}
  \label{fig: groundstate's FzFx}
\end{figure}

Figure~\ref{fig: groundstate's FzFx} shows the $q$-dependence of the longitudinal and transverse magnetizations of the condensate at absolute zero with $n=1\times 10^{10}~\mathrm{cm}^{-3}$. The mean-field result is superimposed for comparison. From Fig.~\ref{fig: groundstate's FzFx}, we find that the phase boundary between the BA and polar phases lies at $q=q_\mathrm{b}=2.05|c_1|n$. The first-order self-consistent approximation discussed in Sec.~\ref{sec:Finite-temperature phase diagram under the Popov approximation} with the same atomic density gives $q_\mathrm{b}=2.02|c_1|n$. These results show that the anomalous average $\tilde{m}^{(\mathrm{R})}_{ij}$ further expands the region of the BA phase from the result of the first-order self-consistent approximation. This can be understood by considering the solution to the coupled equations~(\ref{eq:phi0})-(\ref{eq: number equation 2}) for $q\geq 2|c_1|n$. The lowest-order condensate wave function, which is the solution to Eq.~\eqref{eq:phi0}, is the same as that of the mean-field ground state, and it is given by $\bm{\varphi}^{(0)}=(0,1,0)^\mathrm{T}$ for $q\geq 2|c_1|n$. Since the atoms are condensed in the $i=0$ state, the matrices $\tilde{n}_{ij}$ and $\tilde{m}_{ij}$ are dominated by the matrix elements $\tilde{n}_{00}$ and $\tilde{m}_{00}$, respectively. The higher-order condensate wave function $\bm{\varphi}^{(2)}$, which is the solution to Eq.~\eqref{eq: gGP Castin}, can then be obtained as:
\begin{equation}
\bm{\varphi}^{(2)}=
\begin{cases}
(0,1,0)^\mathrm{T}\ \ \ \ \ \ \ \ \ \  (\mathrm{polar}) & \text{if } \xi\geq 2\\
\begin{pmatrix}
\sqrt{(2-\xi)/8}\\
\sqrt{(2+\xi)/4}\\
\sqrt{(2-\xi)/8}
\end{pmatrix}
\; (\mathrm{BA})& \text{if } \xi<2,
\end{cases} 
\end{equation}
where $\xi\simeq [q-c_0(\tilde{n}_{00}+\tilde{m}^{(\mathrm{R})}_{00})]/(|c_1|n)$. The phase boundary between BA and polar phases is, therefore, given by $\xi=2$, or $q=q_\mathrm{b}\simeq2|c_1|n+c_0(\tilde{n}_{00}+\tilde{m}^{(\mathrm{R})}_{00})$. At absolute zero, $\tilde{n}_{00}$ and $\tilde{m}^{(\mathrm{R})}_{00}$ are both positive (see Sec. \ref{subsec: Renormalized Anomalous Average Term}). Hence, the anomalous average further enhances the shift of the phase boundary toward the polar phase region. Note that the perturbative expansion breaks down in the critical parameter region $2|c_1|n<q<q_\mathrm{b}$ because in this region the system undergoes a quantum phase transition from the polar to the BA phase as the total number density $n$ is increased from zero to the final value [see below Eq.~\eqref{eq:expansion}]. However, the value of $F^\mathrm{c}_\perp/n^\mathrm{c}$ shown in Fig. \ref{fig: groundstate's FzFx} (indicated by the double arrow) is found to be consistent with the expansion of the BA phase region at least qualitatively.
 
\begin{figure}[tbp] % float placement: (h)ere, page (t)op, page (b)ottom, other (p)age
  \centering
  % file name: F:/Paper Drafts (May 17th, 2011)/Figures/phaseboundary.eps
  \includegraphics[bb=0 0 294 224,width=3in,height=2.29in,keepaspectratio]{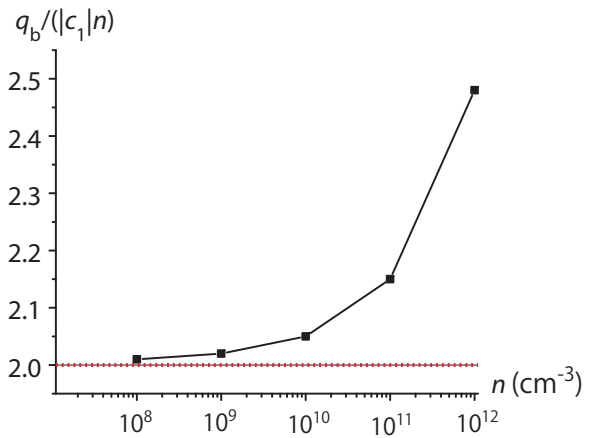}
  \caption{Position $q_\mathrm{b}$ of the phase boundary between the BA and polar phases versus atomic density $n$. The dotted line shows the mean-field value $q_\mathrm{b}=2|c_1|n$. The values of $q_\mathrm{b}$ for $n=10^{11}$ and $10^{12}~\mathrm{cm}^{-3}$, for which the perturbative expansion no longer gives quantitatively reliable results, are plotted only to show their rough estimates.}
  \label{fig:phase boundary}
\end{figure}

Figure~\ref{fig:phase boundary} plots the value of $q_\mathrm{b}$ at the phase boundary between the BA and polar phases for various values of the total number density $n$. We find that the effect of quantum depletion on the spinor order parameter becomes more significant for higher atomic densities, which in turn, leads to a greater expansion of the BA phase from the mean-field result. This trend in the shift of the phase boundary is clearly seen, eventhough the $\chi^{1/2}$ perturbation method no longer gives quantitatively reliable results for atomic density above $n_\mathrm{thres}=10^{10}\ \mathrm{cm}^{-3}$. From these results, we conclude that the quantum depletion significantly alters the mean-field ground-state phase diagram of the spin-1 ferromagnetic BEC. In particular, when the atomic density is larger than $10^{10}~{\rm cm}^{-3}$, which is the case with usual experiments~\cite{Vengalattore08,Vengalattore10}, the system should be treated as a strongly interacting Bose gas. We shall examine this region in a future publication.

%%%%%%%%%%%%%%%%%%%%%%%%%%
\section{Conclusions}
\label{sec:conclusions}
We have studied the interplay between the condensate and noncondensed atoms in a spin-1 $\Rb$ Bose gas in the presence of a quadratic Zeeman effect. First, to investigate the effect of thermal fluctuations on the condensation and magnetism of the system, we have applied the first-order self-consistent approximation and obtained the finite-temperature phase diagram. We find that the system can undergo a two-step phase transition for a certain region of the quadratic Zeeman energy: as the temperature decreases, the thermal gas first enters a nonmagnetized superfluid phase (polar phase), and then a superfluid phase with transverse magnetization (broken-axisymmetry phase). That condensation and spontaneous magnetization occur at different temperatures is characteristic of spinor condensates. Furthermore, via coupling to the magnetized condensate, spin coherence of noncondensed atoms in different magnetic sublevels emerges, leading to magnetization of the noncondensate. By investigating the temperature dependence of magnetization of the noncondensate, we find that the magnetization of the condensate and that of the noncondensate are antiparallel to each other over a broad range of temperatures, except $T\lesssim 0.01~T_0$, where $T_0$ is the transition temperature for a uniform ideal scalar Bose gas with the same atomic density. For $T\lesssim 0.01~T_0$, they become parallel to each other due to quantum depletion. This remarkable feature of the noncondensate at ultralow temperatures makes a distinction from high-temperature atomic thermal clouds.

In contrast to scalar Bose-Einstein condensates (BECs), in spinor BECs the effect of a small fraction of noncondensed atoms on the system's magnetism cannot be ignored. This results from the fact that a large ratio of the spin-independent to spin-dependent interactions can significantly magnify the effect of a small number of noncondensed atoms. To examine the effect of quantum depletion and that of the anomalous average on the magnetism of the system at absolute zero, we have applied the perturbative expansion in powers of $\chi^{1/2}$, where $\chi$ is the noncondensate fraction, to a $\Rb$ spinor Bose gas. From the result, we have found that even a very small noncondensate fraction can make a significant modification of the ground-state phase diagram from the mean-field result. We have also found that when the atomic density exceeds a threshold $n_\mathrm{thres}\sim10^{10}~\mathrm{cm}^{-3}$, the deviation of the order parameter from the mean-field result is so large that the applied perturbation method can no longer give quantitatively reliable results. Therefore, a system with a higher density, which is usually the case with current experiments, should be treated as a strongly interacting spinor Bose gas. This is an interesting subject for future experimental and theoretical studies. However, to deal with this exciting regime, the system must be cooled below temperature $T_\mathrm{qd}$, at which quantum depletion starts to dominates. Although the ratio $T_\mathrm{qd}/T_0$ becomes larger as the atomic density increases, it is generally of the order of $T_\mathrm{qd}/T_0\sim0.01$, which presents a challenge in cooling techniques. 

%%%%%%%%%%%%%%%%%%%%%%%%%%
\begin{acknowledgments}
This work was supported by KAKENHI (22340114, 22740265, 22103005), a Global COE Program \textquotedblleft the Physical Sciences Frontier", and the Photon Frontier Network Program, from MEXT of Japan, and by JSPS and FRST under the Japan-New Zealand Research Cooperative Program. NTP and YK acknowledges P. B. Blakie for useful discussions.
\\
\\
\end{acknowledgments}

%%%%%%%%%%%%%%%%%%%%%%%%%%
\appendix
\section{Derivation of Eq.~(\ref{eq: expectation value identity for Heisenberg equation in Castin})}
\label{sec: Derivation of Eq. expectation value identity for Heisenberg equation in Castin}
The condensate operator $\hat{a}$ and noncondensate operators $\ann{i}(\R)$ can be expressed in terms of the field operator $\annih{i}(\R)$ and the condensate wave function $\varphi_i(\R)$ as
\begin{equation}
\hat{a}=\integral \sum_i\varphi_i^*(\R)\annih{i}(\R),
\label{eq:hat_a}
\end{equation}
\begin{equation}
\ann{i}(\R)= \sum_j \hat{Q}_{ij}\circ \annih{j}(\R) \equiv \int \sum_j \text{d}\R' Q_{ij}(\R,\R') \annih{j}(\R'),
\label{eq:ann_i}
\end{equation}
where $Q_{ij}(\R,\R')=\delta_{ij} \delta(\R-\R')-\varphi_i(\R)\varphi_j^*(\R')$ is the projection operator onto the subspace orthogonal to the condensate wave function $\varphi_i(\R)$. From Eq.~\eqref{eq:ann_i}, we have
\begin{equation}
\integral \sum_i\varphi_i^*(\R)\ann{i}(\R) = 0,
\end{equation}
i.e., the condensate and noncondensate are orthogonal to each other. The commutation relations between the condensate and noncondensate operators are given by
\begin{subequations}
\begin{align}
[\ahat,\adagg]=&1, \\
[\ann{i}(\R),\hat{a}^\dagger]=&0, \\
[\ann{i}(\R),\cre{j}(\R')]=&Q_{ij}(\R,\R'), \\
\text{the others}=&0.
\end{align}
\label{eq:commutation relation}
\end{subequations}

Using Eqs.~(\ref{eq:hat_a}) and (\ref{eq:ann_i}), we obtain
\begin{widetext}
\begin{align}
\langle\hat{a}^\dagger(t)\ann{i}(\R,t)\rangle=&\ \Bigg\langle\left(\int\text{d}\R'\sum_j\varphi_j(\R',t)\creat{j}(\R',t)\right)%\nonumber\\
\left(\integrals \sum_l Q_{il}(\R,\s,t)\annih{l}(\s,t)\right)\Bigg\rangle \nonumber\\
=&\int\text{d}\R'\integrals \sum_{j,l}\varphi_j(\R',t)Q_{il}(\R,\s,t)%\nonumber\\
\underbrace{\langle\creat{j}(\R',t)\annih{l}(\s,t)\rangle}_{\rho^1_{lj}(\s,\R',t)}\nonumber\\
=&\integrals \sum_l Q_{il}(\R,\s,t)%\nonumber\\
\underbrace{\left(\int\,\text{d}\R'\sum_j \rho^1_{lj}(\s,\R',t)\varphi_j(\R',t)\right)}_{N^\mathrm{c}\varphi_l(\s,t)}\nonumber\\
=&\ N^\mathrm{c}\integrals\sum_l Q_{il}(\R,\s,t)\varphi_l(\s,t) \nonumber\\
=&\ 0.
\end{align}
\end{widetext}
Hence, Eq.~(\ref{eq: expectation value identity for Heisenberg equation in Castin}) is proved.

%%%%%%%%%%%%%%%%%%%%%%%%%%%%
\section{Derivation of Eq.~(\ref{eq: order chi 0})}
\label{sec: derivation of eq: order chi 0}
By expanding both sides of Eq.~(\ref{eq: R_i (i=0,...,4)}) up to the order of $\chi^0$, and using Eq.~(\ref{eq: expectation value identity for Heisenberg equation in Castin}), we have
\begin{equation}
\integrals \langle R_0(\R,\s,t) \rangle=0.
\end{equation}
From Eq.~(\ref{eq:expression for R0}) for $R_0(\R,\s,t)$, we obtain
\begin{align}
&\integrals \sum_j Q_{ij}(\R,\s)\Bigg\{\sum_k\Bigg[-i\hbar\delta_{jk}\roundt+(h_0)_{jk}\nonumber\\
&+c_0N\delta_{jk}\sum_l|\varphi_l^{(0)}(\s)|^2\Bigg]\varphi_k^{(0)}(\s,t)\nonumber\\
&+c_1N\sum_{\alpha,k,g,l}(f_\alpha)_{jk}(f_\alpha)_{gl}\varphi_g^{(0)*}(\s)\varphi_{l}^{(0)}(\s)\varphi_{k}^{(0)}(\s)\Bigg\}=0.
\end{align}
From the definition of the projection operator $Q_{ij}(\R,\s)$, we arrive at the time-dependent GP equation:
\begin{align}
&\sum_{j}\left[ -i\hbar\delta_{ij}\frac{\partial}{\partial t}+(h_0)_{ij}+c_0 N\delta_{ij} \sum_k|\varphi_k^{(0)}|^2 \right] \varphi_j^{(0)}(\R,t) \nonumber \\
&+c_1 N \sum_{\alpha,j,k,l}(f_\alpha)_{ij}(f_\alpha)_{kl}\varphi_k^{(0)*}\varphi_{l}^{(0)}\varphi_{j}^{(0)}\nonumber\\
&=\eta^{(0)}(t)\varphi_i^{(0)}(\R,t),
\end{align}
where $\eta^{(0)}(t)$ is an arbitrary real function corresponding to a unitary transformation as described below Eq.~\eqref{eq: order chi 0}. Here, the number of particles in the condensate $N^\mathrm{c}=\langle\adagg\ahat\rangle=N(1-\chi)$ is replaced by the total number of particles $N$ with an error of the order of $\chi^1$, which can be neglected up to the order of $\chi^0$.

%%%%%%%%%%%%%%%%%%%%%%%%%%%%%
\section{Derivation of Eq.~(\ref{eq: order chi 1/2})}
\label{sec: derivation of eq: order chi 1/2}
First, by using Eq.~(\ref{eq: expectation value identity for Heisenberg equation in Castin}) and expanding both sides of Eq.~(\ref{eq: R_i (i=0,...,4)}) up to the order of $\chi^{1/2}$, we obtain
\begin{equation}
\integrals \Big(\langle R_0(\R,\s,t)\rangle+\langle R_1(\R,\s,t)\rangle \Big)=0.
\label{eq: <R0+R1>}
\end{equation}
To calculate the second term in Eq.~(\ref{eq: <R0+R1>}), we use the property of the condensate that the atomic number fluctuation in the condensate is of the order of $\Delta N^\mathrm{c}/N^\mathrm{c}\sim1/\sqrt{N^\mathrm{c}}$. The expectation value $\langle R_1(\R,\s,t)\rangle$ then vanishes because compared with the lowest-order terms in $R_0$, the order of magnitude of terms in $R_1$ is, for example,
\begin{align}
\frac{\langle\adagg\adagg\ahat\ann{i}(\R)\rangle}{\langle\adagg\adagg\ahat\ahat\rangle}=&\frac{\langle\adagg(N^\mathrm{c}+\Delta \hat{N}^\mathrm{c})\ann{i}(\R)\rangle}{\langle\adagg\adagg\ahat\ahat\rangle} \nonumber\\
=&\frac{N^\mathrm{c}}{\langle\adagg\adagg\ahat\ahat\rangle}\underbrace{\langle\adagg\ann{m}(\R)\rangle}_{0}+\frac{\langle\adagg\Delta \hat{N}^\mathrm{c} \ann{i}(\R)\rangle}{\langle\adagg\adagg\ahat\ahat\rangle} \nonumber\\
\sim&\,\mathcal{O}\left(\frac{\chi^{1/2}}{\sqrt{N^\mathrm{c}}}\right).
\end{align}
Here, the number fluctuation operator is defined as $\Delta \hat{N}^\mathrm{c}\equiv \adagg\ahat-\langle\adagg\ahat\rangle=\adagg\ahat-N^\mathrm{c}, (\Delta N^\mathrm{c})^2\equiv\langle(\Delta\hat{N}^\mathrm{c})^2\rangle$, and for a macroscopic number of particles in the condensate $N^\mathrm{c}\sim N\gtrsim 10^6$, the above term can be neglected up to the order of $\chi^{1/2}$ because $1/\sqrt{N^\mathrm{c}}\ll 1$.

Consequently, only the first term $\langle R_0(\R,\s,t)\rangle$ should be retained in Eq.~(\ref{eq: <R0+R1>}) up to this order, and thus the condensate wave function $\varphi_i^{(1)}(\R)$ must satisfy the same equation as $\varphi_i^{(0)}(\R,t)$, i.e., the time-dependent GP equation:
\begin{align}
&\sum_{j}\left[ -i\hbar\delta_{ij}\frac{\partial}{\partial t}+(h_0)_{ij}+c_0 N\delta_{ij} \sum_k|\varphi_k^{(1)}|^2 \right] \varphi_j^{(1)}(\R,t) \nonumber \\
&+c_1 N \sum_{\alpha,j,k,l}(f_\alpha)_{ij}(f_\alpha)_{kl}\varphi_k^{(1)*}\varphi_{l}^{(1)}\varphi_{j}^{(1)}\nonumber\\
&=\eta^{(1)}(t)\varphi_i^{(1)}(\R,t),
\end{align}
With the condition $\lim\limits_{\chi\to 0}\varphi_i^{(1)}(\R,t)=\varphi_i^{(0)}(\R,t)$ and the normalization condition, it can be shown that $\eta^{(1)}(t)=\eta^{(0)}(t)$ and $\varphi_i^{(1)}(\R,t)=\varphi_i^{(0)}(\R,t)$, i.e., the condensate wave function does not change at this order.

Next, the equation of motion for the noncondensate operator at the lowest order $\hat{\Lambda}_i^{(0)}(\R,t)$ is obtained directly by expanding both sides of Eq.~(\ref{eq: R_i (i=0,...,4)}) up to the order of $\chi^{1/2}$:
\begin{align}
i\hbar\frac{d}{dt}\hat{\Lambda}^{(0)}_i(\R,t)=&\frac{1}{\sqrt{N^\mathrm{c}}}i\hbar\frac{d}{dt}\Big(\hat{a}^\dagger(t)\ann{i}(\R,t)\Big) \nonumber\\
=&\underbrace{\frac{1}{\sqrt{N^\mathrm{c}}}\integrals R_0(\R,\s,t)}_{(*1)}\nonumber\\
&+\underbrace{\frac{1}{\sqrt{N^\mathrm{c}}}\integrals R_1(\R,\s,t)}_{(*2)}.
\end{align}

The first term on the right-hand side vanishes because $\varphi_i^{(1)}(\R,t)$ satisfies the ordinary GP equation, so
\begin{align}
(*1)=&\frac{1}{\sqrt{N^\mathrm{c}}}\adagg\ahat \integrals \sum_j Q_{ij}(\R,\s)\Bigg\{\sum_k\Bigg[-i\hbar\delta_{jk}\roundt\nonumber\\
&+(h_0)_{jk}+c_0N\delta_{jk}\sum_l|\varphi_l^{(1)}(\s)|^2\Bigg]\varphi_k^{(1)}(\s,t)\nonumber\\
&+c_1N\sum_{\alpha,k,g,l}(f_\alpha)_{jk}(f_\alpha)_{gl}\varphi_g^{(1)*}(\s)\varphi_{l}^{(1)}(\s)\varphi_{k}^{(1)}(\s)\Bigg\}\nonumber\\
&=0.
\end{align}
The second term can be written as
\begin{widetext}
\begin{align}
(*2)=&\sum_j\left[(h_0)_{ij}+c_0N\delta_{ij}\sum_l|\varphi_l\order{0}(\R)|^2-\eta\order{0}(t)\delta_{ij}\right]\lamb{j}\order{0}(\R)\nonumber\\
&+c_1N\sum_{\alpha,j,k,l}(f_\alpha)_{ij}(f_\alpha)_{kl}\varphi_k^{(0)*}(\R)\varphi_{l}\order{0}(\R)\lamb{j}\order{0}(\R) \nonumber\\
&+\integrals \sum_j Q_{ij}(\R,\s)\Bigg\{c_0N\sum_l\Big[\varphi_l^{(0)*}(\s)\varphi_j\order{0}(\s)\lamb{l}\order{0}(\s)+\varphi_l\order{0}(\s)\varphi_j\order{0}(\s)\lamb{l}^{(0)\dagger}(\s)\Big]\nonumber\\
&+c_1N\sum_{\alpha,k,g,l}(f_\alpha)_{jk}(f_\alpha)_{gl}\Big[\varphi_g^{(0)*}(\s)\varphi_{k}\order{0}(\s)\lamb{l}\order{0}(\s)+\varphi_{l}\order{0}(\s)\varphi_{k}\order{0}(\s)\lamb{g}^{(0)\dagger}(\s)\Big]\Bigg\}.
\end{align}
%\end{widetext}
By separating terms containing $\hat{\Lambda}_j^{(0)}(\R,t)$ from those containing $\hat{\Lambda}_j^{(0)\dagger}(\R,t)$, we obtain the time-dependent BdG equation for the noncondensate operator $\hat{\Lambda}_i^{(0)}(\R,t)$:
\begin{equation}
i\hbar\frac{d}{dt}\hat{\Lambda}_i^{(0)}(\R,t)=A_{ij} \hat{\Lambda}_j^{(0)}(\R,t) + B_{ij} \hat{\Lambda}_j^{(0)\dagger}(\R,t),
\end{equation}
where
%\begin{widetext}
\begin{align}
A_{ij}=&\left[(h_0)_{ij}+c_0 N\delta_{ij} \sum_l |\varphi_l^{(0)}(\R)|^2-\eta\order{0}(t)\delta_{ij}\right]+ c_1 N \sum_{\alpha,k,l}(f_\alpha)_{ij}(f_\alpha)_{kl}\varphi_k^{(0)*}(\R)\varphi_{l}^{(0)}(\R) \nonumber \\
&+ \sum_{k,l}\hat{Q}^{(0)}_{ik}\circ \Bigg\{ c_0 N \varphi_k^{(0)}\varphi_l^{(0)*}+c_1 N \sum_{g,h}(f_\alpha)_{kg}(f_\alpha)_{hl}\varphi_{h}^{(0)*}\varphi_g^{(0)} \Bigg\}\circ \hat{Q}^{(0)}_{lj},\nonumber\\
B_{ij}=&\sum_{k,l}\hat{Q}^{(0)}_{ik}\circ\Bigg\{ c_0 N \varphi_k^{(0)}\varphi_l^{(0)} +c_1 N \sum_{g,h}(f_\alpha)_{kg}(f_\alpha)_{lh}\varphi_{g}^{(0)}\varphi_{h}^{(0)} \Bigg\} \circ \hat{Q}^{(0)*}_{lj}.
\end{align}
%\end{widetext}

%%%%%%%%%%%%%%%%%%%%%%%%%%%%%
\section{Derivation of Eq.~(\ref{eq: order chi 1})}
\label{sec: derivation of eq: order chi 1}
By expanding both sides of Eq.~(\ref{eq: R_i (i=0,...,4)}) up to the order of $\chi^1$, and using Eq.~(\ref{eq: expectation value identity for Heisenberg equation in Castin}), we have
\begin{equation}
\integrals \Big(\langle R_0(\R,\s,t)\rangle+\langle R_1(\R,\s,t) \rangle+\langle R_2(\R,\s,t)\rangle \Big)=0.
\label{eq:R0+R1+R2}
\end{equation}
The first term in Eq.~(\ref{eq:R0+R1+R2}) is given up to the order of $\chi^1$ by
%\begin{widetext}
\begin{align}
\integrals \langle R_0(\R,\s,t) \rangle=&\integrals \sum_j Q_{ij}(\R,\s)\Bigg\{\sum_k\Bigg[N^\mathrm{c}\left(-i\hbar\delta_{jk}\roundt +(h_0)_{jk}\right) \nonumber\\
&+c_0\Big(\langle(\hat{N}^\mathrm{c})^2\rangle-N^\mathrm{c}\Big)\delta_{jk}\sum_l|\varphi_l\order{2}(\s)|^2\Bigg]\varphi_k\order{2}(\s) \nonumber\\
&+c_1\Big(\langle(\hat{N}^\mathrm{c})^2\rangle-N^\mathrm{c}\Big)\sum_{\alpha,k,g,l}(f_\alpha)_{jk}(f_\alpha)_{gl}\varphi_g^{(2)*}(\s)\varphi_{l}\order{2}(\s)\varphi_{k}\order{2}(\s)\Bigg\}.
\label{eq:R0}
\end{align}
%\end{widetext}
The number fluctuation in the condensate satisfies $(\Delta N^\mathrm{c}/N^\mathrm{c})^2\sim 1/N^\mathrm{c}\ll \chi^1$ (for $N\gtrsim 10^6$), and can be neglected up to this order. Equation~\eqref{eq:R0} then reduces to
%\begin{widetext}
\begin{align}
\integrals \langle R_0(\R,\s,t) \rangle=&N^\mathrm{c}\integrals \sum_j Q_{ij}(\R,\s)\Bigg\{\sum_k\Bigg[-i\hbar\delta_{jk}\roundt +(h_0)_{jk}+c_0N^\mathrm{c}\delta_{jk}\sum_l|\varphi_l\order{2}|^2\Bigg]\varphi_k\order{2}(\s) \nonumber\\
&+c_1N^\mathrm{c}\sum_{\alpha,k,g,l}(f_\alpha)_{jk}(f_\alpha)_{gl}\varphi_g^{(2)*}(\s)\varphi_{l}\order{2}(\s)\varphi_{k}\order{2}(\s)\Bigg\}.
\end{align}
%\end{widetext}
Using the fact that $\Delta N^\mathrm{c}/N^\mathrm{c} \sim 1/\sqrt{N^\mathrm{c}}\ll \chi^{1/2}$, the second term in Eq.~(\ref{eq:R0+R1+R2}) vanishes up to this order:
\begin{align}
\integrals \langle R_1(\R,\s,t) \rangle \sim \, \mathcal{O}\left(\frac{\chi^{1/2}}{\sqrt{N^\mathrm{c}}}\right)\ll \chi^1.
\end{align}
The third term in Eq.~(\ref{eq:R0+R1+R2}) can be rewritten as
\begin{align}
\integrals \langle R_2(\R,\s,t)\rangle=&N^\mathrm{c}\integrals\sum_j Q_{ij}(\R,\s)\Bigg\{c_0\sum_l\Big[\tilde{n}_{ll}(\s)\varphi_j\order{2}(\s)+\tilde{n}_{jl}^*(\s,\s)\varphi_l\order{2}(\s)+\tilde{m}_{jl}(\s,\s)\varphi_l^{(2)*}(\s)\Big] \nonumber\\
&+c_1\sum_{\alpha,k,g,l}(f_\alpha)_{jk}(f_\alpha)_{gl}\Big[\tilde{n}_{gl}(\s,\s)\varphi_{k}(\s)+\tilde{n}_{kg}^*(\s,\s)\varphi_{l}(\s)+\tilde{m}_{kl}(\s,\s)\varphi_g^{(2)*}(\s)\Big]\Bigg\}-\mathcal{F}_i(\R),
\end{align}
where $\mathcal{F}_i(\R)$ is defined as
\begin{align}
\mathcal{F}_i(\R)\equiv& \integrals \Bigg\{ c_0 N^\mathrm{c} \left(\sum_l|\varphi\order{0}_l(\s)|^2 \right)\sum_j\Big[ \tilde{n}_{ij}^*(\R,\s)\varphi\order{0}_j(\s) + \tilde{m}_{ij}(\R,\s)\varphi^{(0)*}_j(\s) \Big] \nonumber\\
&+c_1 N^\mathrm{c} \sum_{\alpha,j,k,g,l}(f_\alpha)_{jk}(f_\alpha)_{gl}\varphi^{(0)*}_g(\s)\varphi\order{0}_{l}(\s)\Big[\tilde{n}_{ij}^*(\R,\s)\varphi\order{0}_{k}(\s) + \tilde{m}_{ik}(\R,\s)\varphi^{(0)*}_j(\s)\Big] \Bigg\}.
\end{align}
\end{widetext}
By substituting the above expressions into Eq.~(\ref{eq:R0+R1+R2}), and using the definition of $Q_{ij}(\R,\s)$ in Eq.~\eqref{eq:ann_i}, we obtain the generalized GP equation for $\varphi\order{2}_i(\R)$ given in Eq.~(\ref{eq: order chi 1}). Note that $\mathcal{F}_i(\R)$ does not change under the operation of $Q_{ij}(\R,\s)$, i.e.,
\begin{align}
\integrals \sum_j Q_{ij}(\R,\s)\mathcal{F}_j(\s)=\mathcal{F}_i(\R).
\end{align}

%%%%%%%%%%%%%%%%%%%%%%%%%%

\end{document}